\documentclass[useAMS,usenatbib]{mn2e}

\usepackage{amsmath,graphicx,lipsum,overpic,amssymb}

\usepackage{subfigure}

\newcommand{\ignore}[1]{}

\def\mdot{\dot{M}}

\title[ Applications of Planet Formation to the HL Tau System ]{ Physics of Planet Trapping with Applications to HL Tau }
\author[A.J. Cridland, R.E. Pudritz, \& M. Alessi]{A.J. Cridland$^{1,2}$\thanks{E-mail: cridland@strw.leidenuniv.nl}, Ralph E. Pudritz$^{2,3}$\thanks{E-mail: pudritz@mcmaster.ca}, \& Matthew Alessi$^{2}$\\
$^1$Leiden Observatory, Leiden University, 2300 RA Leiden, the Netherlands \\ $^{2}$Department of Physics and Astronomy, McMaster University, Hamilton, Ontario, Canada, L8S 4E8 \\ $^3$Origins Institute, McMaster University, Hamilton, Ontario, Canada, L8S 4E8}

\begin{document}
\bibliographystyle{mn2e}
\date{\today}

\pagerange{\pageref{firstpage}--\pageref{lastpage}} \pubyear{2018}

\maketitle

\label{firstpage}

\begin{abstract}
We explore planet formation in the HL Tau disk and possible origins of the prominent gaps and rings observed by ALMA. We investigate whether dust gaps are caused by dynamically trapped planetary embryos at the ice lines of abundant volatiles. The global properties of the HL Tau disk (total mass, size) at its current age are used to constrain an evolving analytic disk model describing its temperature and density profiles. By performing a detailed analysis of the planet-disk interaction for a planet near the water ice line including a rigorous treatment of the dust opacity, we confirm that water is sufficiently abundant ($1.5\times 10^{-4}$ molecules per H) to trap planets at its ice line due to an opacity transition. When the abundance of water is reduced by 50$\%$ planet trapping disappears. We extend our analysis to other planet traps: the heat transition, dead zone edge, and the CO$_2$ ice line and find similar trapping. The formation of planets via planetesimal accretion is computed for dynamically trapped embryos at the water ice line, dead zone, and heat transition. The end products orbit in the inner disk ($R<3$ AU), unresolved by ALMA, with masses that range between sub-Earth to 5 Jupiter masses. While we find that the dust gaps correspond well with the radial positions of the CO$_2$, CH$_4$, and CO ice lines, the planetesimal accretion rates at these radii are too small to build large embryos within 1 Myr. 
\end{abstract}


\section{ Introduction }\label{sec:intro4}

{\it What process or processes are responsible for the gaps and rings structure in disks like HL Tau?}   A number of explanations have been proposed including the existence gap-opening planets \citep{Dipierro2015,Tamayo15}, efficient grain growth \citep{Pinte2016}, and sintering-induced grain fragmentation near ice lines \citep{Zhang15,Okuzumi2016}. Rings of bright emission and the gaps between them are not restricted to HL Tau, and appear to be a common feature in protoplanetary disks \citep{Zhang16}.   

The correspondence between the gaps in the HL Tau disk and volatile ice lines (also known as condensation fronts) has been studied previously by \cite{Zhang15}. They used a temperature profile derived from a 2D radiative transfer model and assumed that ice lines would appear at the radii where the gas reached the sublimation temperature for a number of abundant volatiles. \cite{Zhang16} go on to suggest a connection between ice lines, dust trapping, and emission gaps. Because dust trapping is an important first step in the process of planet formation, we might expect that the first planetary seeds (often called planetary embryos), or up to fully formed planets may also be found in conjunction with the radial position of ice lines in these young systems.      

The nature of this connection must take into account the fact that planets exchange angular momentum with their host disks and will migrate through them \citep{GoldTrem79,Ward1997,Paard10}. Linear solutions of the total torque on small ($\sim$ M$_\oplus$) planets suggest that their migration will occur in only $ 10^5 $ yr which is a small fraction of the disk lifetime, potentially threatening the survivability of planets in disks \citep{Masset06}. This well known rapid Type I migration problem of low mass planetary cores arises in simplified disk models wherein various kinds of expected inhomogeneities - such as ice lines - are ignored.  Inhomogeneities, however,  can create localized planet traps - null points in the net torque - where rapid migration is arrested \citep{Lyra2010,Horn2012,HP11,HP13,D14,Cole14} thereby solving this rapid migration problem.   Planets, trapped in such regions, will move along with them at rates closer to the viscous evolution time of the disk until they achieve masses of $\sim 10$ Earth masses at which point they are liberated from their trap through the saturation of the corotation torque (see \cite{Crid16a}) or through opening an annular gap in the disk. Both the saturation of the corotation torque and gap opening depend on the local gas viscosity, and hence on the ionization state of the gas. In this way,  planet formation becomes linked to the position and movement of traps, as well as their microphysics over the full lifetime of the disk. This has been previously been shown by both analytic and numerical studies (see review by \cite{Pudritz2018}).
 
This raises two important questions about the nature of planet building and disk features in the HL Tau system.  (1) What chemical species have ice lines that are capable of trapping planets?   (2) Can  planets also appear at the other traps?   We note that while it has already been suggested that the water ice line acts as a trap \citep{Lyra2010}, do other species provide traps as well (eg. CO or CO$_2$ ice lines)?  Those that can, should harbour growing planetary embryos, which could create pressure maxima or even outright gaps in the disks.   Those that cannot serve as traps would then be expected to only provide ring features arising from opacity transitions \citep{Zhang16}.  The answers to these questions require that we are able to locate the positions of predicted traps at the particular age of an observed star-disk system.  

In this paper we address these two problems, specifically; (1)  what species are capable of trapping planets at their associated ice line radii in evolving disks, and (2) using our combined chemical and planet formation models where are the traps in the observed HL Tau disk and what mass planets - if any - might they harbour?   

The water ice line is the archetype for planet trapping and works because its opacity transition modifies the disk's temperature profile \citep{Lyra2010,D14}. Thus to address the first problem - what species can produce planet traps - we construct a general physical theory which analyzes the opacity transition across the water ice line by computing both the radial distribution of the ice from a photo-chemical disk model and the optical properties of an ice-silicate grain mixture. We generalize this theory to study planet trapping at the ice lines of other abundant volatiles like CO and CO$_2$.

To accomplish these goals, we first generate HL Tau disk models that reproduce the global properties (mass, size, age) of the HL Tau disk. We tune our initial conditions so that after roughly 1 Myr of disk evolution one recovers the observed properties (ie. gas and dust mass, radial size) of the HL Tau disk. Then, using a complex photo-chemical code we compute the radial distribution of abundant volatiles (in their ice and gas phases) and find coincidence between the radial location of their ice lines with the observed gaps. We can also compute the growth of planetary embryos that can form at the ice line of abundant volatiles and elucidate the details of Type-I migration around the opacity transition induced by an ice line. These embryos would perturb the gas creating pressure maxima towards which the dust will flow, resulting in the observed gaps in emission \citep{Dipierro2015}.   

To address our second question - what planets might have formed in the current HL Tau disk - we investigate planet formation in HL Tau by applying models that we have developed in \cite{Crid16a,Crid16b,Crid17} to our HL Tau disk models. This planet formation model combines the standard planetesimal accretion model \citep{KI02,IL04a} with a planet trap model to limit the inward migration rate of the proto-planet \citep{Masset06,HP13} within the predicted disk age of 1 Myr. We also generalize our analysis of planet trapping to the other planet traps (heat transition, \cite{Lyra2010,Horn2012} and dead zone, \cite{MP09,RSC2013}) that have appeared in our previous formation models. 

We derive the radial locations of ice lines by computing the photo-chemistry of our evolving analytic disk model, and compare their locations to the radial locations of the gaps observed today, and to the results of \cite{Zhang15}. We define the location of the ice line to be the point where the ice abundnace exceeds the abundance of the corresonponding vapour. This definition has served well in the past in thermochemical equilibrium calculations (eg. \cite{APC16a}) where the transition from vapour to ice is very sharp. It is important to note that in light of our more complete treatment of the astrochemistry, the transition between water vapour and ice can be (radially) slow, and in the case of water spans roughly 4 AU in our model (see Figure \ref{fig:p4:new01x}). In this case a better definition for the `location' of the ice line would be where the abundance of the volatile ice exceeds half of its total abundance.

We note that models of `pebble accretion' show that the growth of a planetary core may be dominated by the accretion of cm-sized `pebbles' rather than 10-100 km sized planetesimals \citep{LambJoh2014,Bitsch2015}. Pebble accretion as a formation mechanism for rocky planetary cores is an active field of research.  However, recent numerical studies  have suggested that the direct accretion of pebbles onto a planetary core is limited to only $\sim 1$ M$_\oplus$ because the pebbles are ablated in the planetary envelope \citep{Brouwers2018} and are then recycled back into the surrounding protoplanetary disk \citep{Alibert2017}. The efficiency of this recycling is not well constrained by hydrodynamic simulations. In 2D and 3D simulations \cite{Ormel2015} and \cite{Ormel2015b} find that only 1\% of the envelope mass per orbital time is filtered back into the disk, while \cite{LamLega2017} and \cite{Fung2015} find 10\% and 100\% respectively. Given these current uncertainties, our work continues for now, to focus on planetesimal accretion for the growth of planetary embryos.   

The paper is organized as follows: in \S\ref{sec:back4} we outline background information relevant to this paper and outline the disk parameters selected to mimic global properties of the HL Tau disk. In \S\ref{sec:gapsandtraps} we compare the radial location of the ice lines in the disk to the locations of the dust gaps observed by ALMA and analyse their potential for hosting trapped planets. In \S\ref{sec:GlobTrap} we generalize our analysis to compute the details of planet trapping at the heat transition and dead zone planet traps. In \S\ref{sec:resPlnt4} we apply our model to the HL Tau system and report the results of our planet formation model for a range of disk parameters which cover the observed range of gas mass in the disk around HL Tau. In \S\ref{sec:con4} we outline our results discuss the relevance and limitations of this work.

\section{ Model Background }\label{sec:back4}

Our theoretical framework is an `end-to-end' analysis that links the structure and evolution of an accretion disk down to the formation and growth of a young planet. This framework combines an analytic gas disk model, a numerical dust physics model, a Monte Carlo radiative transfer scheme, a time-dependent astrochemical code, and a planet formation model to ultimately compute the chemical composition of the gas that is delivered to the atmosphere of a growing planet (for technical details see \cite{Crid16a} and \cite{Crid16b}). 

A similar method has been developed by \cite{Mordasini16} (and reviewed in \cite{Mordasini18}) which they call a `chain' model. The chemical evolution of the gas is not computed in their model, and volatile abundances were either prescribed by observational data, or removed entirely. In this way, the majority of elements heavier than helium were accreted onto planets as ice frozen to planetsimals. \cite{Mordasini16} do, however, compute the final equilibrium structure and molecular abundances of the resulting planetary atmosphere - something that is beyond the scope of this work.

In our framework, the models that govern physics and chemistry are computed in succession. The ordering of our calculations begins with the prescription of a gas model for an evolving disk. The following steps are: {\it gas disk model} $\rightarrow$ {\it dust model (radial drift, growth / fragmentation)} $\rightarrow$ {\it UV / X-Ray radiative transfer} $\rightarrow$ {\it time dependent chemistry} $\rightarrow$ {\it planetesimal accretion planet formation}. Each step, much as in the chain model,  depends on the preceding ones, and impact the results of all following ones. In doing this we can successively model complicated interactions like the impact of dust evolution on the radiative transfer of high energy radiation (see \cite{Crid16b}). 

Any particular model in this scheme acts as an input for all subsequent calculations, while it has no impact on models that it follows. One such example is the existence and evolution of the turbulent dead zone (which depends on the ionization, computed in the chemistry step) should, in principle impact the settling and fragmentation of large dust grains. However the ionization depends on the flux of ionizing radiation, and hence on the distribution of the dust grains. This could in principle be solved by iterating the {\it dust model} $\rightarrow$ {\it rad transfer} $\rightarrow$ {\it chemistry} steps, however this is computationally expensive. Additionally we showed in \cite{Crid16b} that such a calculation has a minimal impact on the radial extent of the dead zone.

\subsection{Gas disk model}

The details of our model are outlined in \cite{Crid16a}, \cite{Crid16b}, and \cite{APC16a}, and we refer readers to those papers for the technical details. Here we will summarize some of its most important features. The gas disk model is a self-similar, analytic model as seen in \cite{Cham09}. It computes the co-evolving radial structure of the gas surface density and midplane temperature as mass viscously accretes through the disk. The radial structure follows a power law of the form: \begin{align}
\Sigma(r,t)\propto \Sigma_0(t)r^{-p}\quad T(r,t)\propto T_0(t)r^{-q},
\label{eq:p401}
\end{align}
where the exponents $p$ and $q$ are determined by the source of gas heating and the time dependent functions $\Sigma_0$ and $T_0$ depend on the mass accretion rate, and hence on our choice of $\alpha$ in the standard $\alpha$-disk prescription of the disk viscosity \citep{SS73}. In what follows we assume a disk $\alpha=10^{-3}$, their functional form can be found in \cite{Cham09}.

We note here that historically $\alpha$ was used to describe the strength of gas turbulence in the disk - leading to momentum transport and hence accretion. However recently (in light of the existence of dead zones, see below) $\alpha$ has been used simply as a parameterization of the rate of accretion through the disk. In past work (see \cite{Crid17} and \cite{APC16a}) we assume two components of angular momentum transport: turbulence and magnetic disk winds - each contributing to maintain a constant (in time and space) `disk $\alpha$': \begin{align}
\alpha = \alpha_{\rm turb} + \alpha_{\rm wind}.
\end{align}
Such a distinction becomes important when including the effects of a turbulent dead zone on the dynamics of the dust disk (discussed below).

There are two primary heating sources in an accretion disk, viscosity and direct irradiation. The former is ultimately caused by the conversion of  gravitational potential energy into heat as the gas accretes through the viscous disk, while the latter is caused by the absorption and re-emission of stellar radiation by the upper atmosphere of the disk. Generally speaking, viscous heating dominates the inner region of the disk, while direct irradiation becomes important at larger radii. 

At a certain radius, the radiative heating becomes more important than the heating from viscous evolution. We call this radius the heat transition ($r_t$), so generally the surface density and temperature of the gas is given by:\begin{align}
\frac{\Sigma}{\Sigma_0}(r) &\propto \begin{cases}
r^{-3/5} & r < r_t \\
r^{-15/14} & r > r_t
\end{cases} \label{eq:p402x}\\
\frac{T}{T_0}(r) &\propto \begin{cases}
r^{-9/10} & r < r_t \\
r^{-3/7} & r > r_t
\end{cases}.
\label{eq:p402}
\end{align}
As the mass accretion rate drops, viscous heating is less efficient and $r_t$ moves inward. Eventually the temperature structure is fully determined by direct irradiation, similar to the results of \cite{CG97}.

In light of our above assumption that the effective $\alpha$ associated with angular momentum transport can be generated by both turbulence or magnetic winds, we note that the effective viscosity responsible for heating in the inner disk is generated by both the Reynolds (turbulent) stress and Maxwell (magnetic) stress. Hence upon entering the dead zone (where $\alpha_{\rm turb}$ falls by at least two orders of magnitude), heating is dominated by non-ideal MHD effects (eg. ambipolar diffusion heating) which, in steady state theory heats the disk at a rate that scales with the release of gravitational potential energy \citep{CieslaCuzzi2006,BaiStone2013}. 

We assume that the disk $\alpha$ stays constant throughout the disk, and that the functional form of the viscosity ($\nu =  \alpha c_s H$) remains the same regardless of the primary source of the viscosity. Such an assumption seems reasonable because when the magnetic field is in equipartition, the Alfv\`en speed scales with the local sounds speed ($c_s$) and the relevant length scale will still be related to the gas scale height ($H$). Then the effective $\alpha$ is related to the ratio of the magnetic pressure to the gas pressure, as well as the local orientation of the magnetic field \citep{BaiStone2013}.

One drawback of using an effective viscosity to handle both turbulent and MHD disk wind torques is that while the former  leads to radial outward spreading of the disk, winds transport carries away angular momentum leading to shrinkage of the disk \citep{PN86}.  Indeed, in recent models that include non-linear MHD effects, the spread of the disk is suppressed \citep{Bai2016}.  This probably will not effect he dynamics in the inner, planetary forming regions of the disk.

\subsection{Dust model}

The dust is allowed to evolve separately from the gas, and its evolution depends on the level of coupling between a dust grain and the gas. The parameter which describes this coupling is known as the Stokes number - the ratio of the grain stopping time to eddy turnover time:\begin{align}
St = \frac{a\rho_s}{\Sigma}\frac{\pi}{2},
\label{eq:p403}
\end{align}
where $a$ is the radius of the grain, $\rho_s$ is the internal density of the grain, and $\Sigma$ is the gas surface density. When $St<1$ the grain is well coupled to the gas and will primarily evolve with the viscous evolution of the gas. While when $St>1$ the dust will decouple from the gas and orbit the host star in Keplarian orbits. Because the gas is pressure supported and orbits at sub-Keplarian speeds the large grains will feel a head wind as they pass through the gas. This head wind drains angular momentum and moves the grain into smaller orbits. This process is known as radial drift \citep{W77}, and proceeds on a time scale of: \begin{align}
\tau_{drift} = \frac{rV_k}{St c^2_s}\gamma^{-1},
\label{eq:p404}
\end{align}
where $V_k$ is the Kepler speed, $c_s$ is the gas sound speed, and $\gamma$ is the absolute magnitude of the gas pressure gradient. Radial drift plays an important role in our model because it impacts both the radial structure of the dust, which changes the flux of high energy radiation \citep{Crid16b}, and provides ample solid material for the formation of planetesimals at small radii \citep{Drazkowska2016,Crid17}. It does, however, clear out the solid material at large radii, and makes planet formation through the accretion of planetesimals more difficult at radii outward of the water ice line.

Along with its radial drift, the dust can coagulate, fragment, and settle. Each of these processes depend sensitively on the level of turbulence in the gas (see for example \cite{DD05}, \cite{B09}, and \cite{B12}) which we assume is a constant $\alpha_{\rm turb} = 10^{-3}$ in most of our calculations. Such an assumption relies on a consistent source of turbulence that is often attributed to the magnetorotational instability (MRI) \citep{BalbusHawley91}. 

Non-ideal magnetohydrodynamic (MHD) calculations \citep{BS11,Simon13,BaiStone2013,BaiStone2016} have suggested that along the midplane of the disk the MRI is suppressed by either Ohmic diffusion or the Hall effect - often called a `dead zone'. In this laminar flow, angular momentum transport is dominated instead by a launching of a magnetize wind \citep{PPV07,BS11,Simon13}. We stress that while the turbulent component of the angular momentum transport may drop within the dead zone, the MHD disk wind maintains the transport of angular momentum required to remain in a steady state disk solution (constant $\alpha = 10^{-3}$).

For the disk models discussed in \S \ref{sec:gapsandtraps} we assume a constant $\alpha_{\rm turb}$ when computing the evolution of the dust. However in \S \ref{sec:GlobTrap}, where the goal is to demonstrate the physics of planet trapping at the dead zone edge we will vary $\alpha_{\rm turb}$ according to the method outlined in \cite{Crid16b} (see also Appendix \ref{sec:app02}).

Within the dead zone of varied turbulent alpha model the gas ionization (as computed by our chemical model discussed below) is too low for turbulence to be efficiently generated by the MRI \citep{BS11}. With the only source of turbulence generated through hydrodynamical sources, the turbulent $\alpha_{\rm turb}$ drops significantly. Here we assume that within the dead zone $\alpha_{\rm turb} = 10^{-5}$ while outward of the dead zone edge $\alpha_{\rm turb} = 10^{-3}$. With the reduction in turbulent power the relative speed of the dust grains is reduced. This reduces the efficiency of fragmentation, which leads to larger grains that can more effectively settle to the midplane of the disk. This is due to both the grains being larger (heavier) and the reduction of turbulence that would otherwise mix grains back up above the midplane.

\subsection{Ionizing radiation}

We compute the transfer of UV and X-Ray radiation because of its importance on the astrochemistry in protoplanetary disks \citep{Fogel11}. We use the Monte Carlo radiative transfer scheme {\it RADMC3D} \citep{RADMC} to compute the flux of photons through out the disk. To properly model the flux from a T Tauri star we compute the radiative transfer for four wavelengths within a range of 930-2000 ${\rm \AA}$ for UV and 1-20 keV for X-Rays. We then extrapolate the remaining wavelengths within the above ranges using a sample spectrum as seen in \cite{Fogel11} and \cite{BB11}. These sample spectra have been normalized to match the total flux from TW Hya, and can be scaled if needed.

\subsection{Disk astrochemistry }

The astrochemistry is computed with a time-dependent gas chemistry code as featured in \cite{Fogel11}, \cite{Cleeves14}, and \cite{Crid16a}. Generally it computes the reaction rates of many ($\sim 5900$) chemical reactions including gas phase, volatile absorption/desorption on grains, grain surface reactions (for the production of H$_2$O and H$_2$), and photo-chemistry. We allow for the chemical structure of the disk to evolve with time by selecting a number of `snapshots' of the gas surface density and temperature. Each snapshot is spaced out in time by $\sim 17000$ yr, totally 300 snapshots over a typical $\sim 4$ Myr disk lifetime. The length of the timestep is about an order of magnitude shorted than the typical viscous time $\sim 1$ Myr in an effort to minimize the effect of gas transport into and out of a particular cell during the chemistry calculation.

Each snapshot includes gas density and temperature profiles, a dust density profile, and the UV and X-Ray radiation fields. As the disk ages, each of these profiles change from one snapshot to the next. Each snapshot begins with the same molecular initial condition and is run until a steady-state solution is found. This method allows the chemistry to be computed in parallel over many snapshots simultaneously, which dramatically reduces the required computing time.

The chemical model is initialized with molecular abundances based on observations of the volatile component of molecular clouds. Of particular importance, water is initialized as an ice with an abundance of 1.5$\times 10^{-4}$ molecules per H-atom. This abundance is used as our fiducial water abundance model used below. While also assuming that the dust follows the typical gas-to-dust mass ratio of 100 we find a maximum water ice-to-silicate mass ratio of about 0.13 outward of the water ice line and approximately zero inward of the water ice line. 

Within the ice line the water ice smoothly transitions between these two mass ratios. Our fiducial mass ratio resides at the lower end of the ice mass ratio inferred from the comet 67P/C-G by \cite{Davidsson2016}. Similarly it is lower than the three test cases of \cite{Bitsch2016}\footnote{the lowest of which was a 0.33 ice-to-silicate mass ratio} who computed a similar test of the impact of the water ice abundance on planet migration. Our choice of water abundance is based on general knowledge of the global astrochemical properties of protoplanetary disks.

The reaction rates that are used to compute the adsorption and desorption of volatiles are based on lab measurements of ice binding strengths, and hence have associated uncertainties. We have not propagated these uncertainties through our calculation of the ice lines radial location (see below), and hence we can not estimate the uncertainties in the radial location of the ice lines. Additionally we assume that all ice is pure when it desorbs - which ignores the desorption of clathrate hydrates (as explored by \cite{Zhang15}), mixes of water ice with trapped gaseous molecules. These species of water ice generally have lower sublimation temperatures than pure water ice, and hence could produce additional ice lines or shift the location of the ice lines that are computed below.

\subsection{Planet formation and migration}

Our planet formation model follows the standard planetesimal accretion model \citep{KI02,IL04a}, where the core of a young planet is built by the successive accretion of $\sim$10-100 km sized objects. Core growth is seeded with a mass of 0.01 M$_\oplus$, and is largely insensitive to the initial embryo mass because the accretion timescale scale with the mass of the core \citep{IL04a}. 

In these semi-analytic models, accreting protoplanets are assumed to grow from planetesimals within a feeding zone of approximately 10 Hill radii ($r_H = a (M_p/3M_*)^{1/3}$). For a planetesimal to accrete its speed relative to the protoplanet should be: $\sigma / \Omega \sim 10 r_H$ \citep{IL04a}. However this prescription generally ignores the combined impact of migration and gravitational scattering of planetesimals by the protoplanet (for ex. \cite{TanakaIda1999}). In practice, the accretion of planetesimals within the feeding zone may not be completely efficient. The effect of this inefficiency is beyond the scope of this work, and we assume that all planetesimals within the feeding zone are available for accretion.

We combine this accretion model with the dynamical trapping of forming planets in `planet traps' to limit the rate of Type-I migration. Planet traps slow the Type-I migration rate of proto-planets because of a combination of the Lindblad and co-rotation torques which can reverse the direction of angular momentum transport \citep{Masset06,HP11}. 

Specifically, planet traps are discontinuities in the physical properties of the disk which causes the direction of the net torque acting on the planet to reverse direction. In past works, the three planet traps which we focussed on were the water ice line, the heat transition, and the outer edge of the dead zone. The discontinuities associated with these traps are the dust opacity, gas temperature profile, and gas turbulence strength respectively. In this work we model these discontinuities and compute the net torque on a planet at each of these traps and demonstrate their influence on planetary migration.

We note that, for simplicity, the models for these discontinuities (ie. transition in turbulent strength) have not been accounted for in computing the radial structure of the gas and dust. In most cases, these discontinuities change the structure of the disk locally, near the planet trap, and do not have an impact on the global structure and overall evolution of our disk model. In past works we have verified that a change in the turbulent strength does not drastically change the global distribution of the dust which act as a primary source of opacity for high energy photons (see the appendix of \cite{Crid16b}). As a result the ionization structure, and hence the size and evolution of the dead zone is not changed when its physical effect on the disk is included in our model.

In our model we assume that the planet is trapped at one of the planet traps, and migrates only because the location of the planet trap changes as the disk ages. This assumption remains valid as long as the co-rotation torque remains unsaturated \citep{Paard10,HP13}. The saturation of the co-rotation torque occurs mainly when the gas turbulence is weak, and so at each timestep we check the saturation of the co-rotation torque if the planet trap is found within the dead zone (see below).

If the co-rotation torque saturates, the planet migrates due to the standard Lindblad torque \cite{Paard10}:\begin{align}
\Gamma = C_\Gamma \Sigma \Omega^2_p r_p^4 \left(\frac{M_p}{M_*}\right)^2\left(\frac{r_p}{|\Delta r|}\right)^2,
\label{eq:p405x}
\end{align}
where $r_p$, $\Omega_p$, $M_p$, and $M_*$ are the radial position, Keplar frequency, mass of the planet, and host star mass respectively. $|\Delta r|$ is the relavent length scale for the density gradient, which we set as the maximum of the disk scale height ($H$) or the Hill radius ($R_H$). The radial evolution of the proto-planet is $\dot{r}_p/r_p = \Gamma/J_p$ such that:\begin{align}
\frac{1}{r_p}\frac{dr_p}{dt} &= -\frac{C_\Gamma \Sigma \Omega_p r_p^2}{M_p} \left(\frac{M_p}{M_*}\right)^2\left(\frac{r_p}{max(H,R_H)}\right)^2,
\label{eq:p405}
\end{align}
The numerical constant $C_\Gamma$ is related to the strength of the net torques on the planet (ie. see \cite{Paard10}).

When the planet is sufficiently massive (few tens of M$_{\oplus}$), its gravitational tug on the surrounding gas will overpower the viscous torques and open a gap in the gas disk. At this point Type-I migration is suppressed and the planet begins Type-II migrating. In this regime, the planet migrations inward on the viscous timescale, as it acts as an intermediary to angular momentum transport through the disk. In this case the orbital radii of the planet evolves as:\begin{align}
\frac{1}{r_p}\frac{dr_p}{dt} = -\frac{\nu}{r_p^2},
\label{eq:p406}
\end{align}
where $\nu$ is the gas viscosity, and $\nu = \alpha c_s H$ for the standard `$\alpha$-disk' \citep{SS73}.

\subsection{ Verifying the process of planet trapping }

Our previous analysis of co-rotation torque saturation assumed that the process was discrete, so that the co-rotation torque saturates only after the planet reached a particular mass (computed by comparing the eddy-turnover rate and libration rate, see \cite{Crid16a}). However \cite{Paard2011} showed that this process actually occurs as a smooth transition between the saturation states of the co-rotation torque. This smooth transition is captured by a set of three numerically motivated functions that weaken the co-rotation torque in regions of the disk where we would have otherwise assumed that it was fully unsaturated. The full form of the total torques that were derived by \cite{Paard2011} was used by \cite{Cole14} to compute the global rate of Type-I migration over their entire disk and across a wide range of planetary masses. In \S\ref{sec:resTrap} we follow the method of \cite{Cole14} to compute the trapping of planets near volatile ice lines, the heat transition, and the dead zone edge.

One purpose of this paper is to verify that planet trapping proceeds near the designated planet traps that are outlined in previous works \citep{Crid16a,APC16a}. This is done by computing the total torque on a planet as was done in \cite{Cole14}, and requires a modification to our original disk model that captures the physics responsible for planet trapping. 

In the case of the ice line this requires a modified disk model where the dust opacity is allowed to change as a function of radius across the ice line. For this model, we assume that the bulk of the solid mass is made up of only 0.1 $\mu$m sized grains with an ice mantle that varies with radius. Hence we take the radial distribution of water ice from previous astrochemical results and compute the resulting optical properties of the dust-ice mixture. The details and results of these calculations are outlined in Appendix \ref{sec:app01} and \S \ref{sec:resTrap} respectively. The purpose of this calculation is to show how a change in the dust opacity due to changing water ice abundance leads to a reversal of the net torque near the ice line.

We expect that trapping at the dead zone is related to the distribution and size of dust grains which impact the opacity, and hence temperature profile of the gas. This assumes the gas and dust temperatures are equal, which is the case at for high densities along the midplane. To test whether the dead zone acts as a trap we compute the density and size distribution of the dust using the numerical model of \cite{B12}, then compute the temperature profile of this dust population using the Monte Carlo code {\it RADMC} \citep{RADMC}. In computing the radiative transfer we are forced to neglect the radial distribution of water ice in order to keep the calculation tractable. We assume that the solid component is made up of (water) icy grains everywhere. In this case we focus primarily on the physical processes that result in planet trapping near the dead zone.

We combine this radiative temperature profile with the viscous temperature profile that we compute analytically to get the total gas temperature profile by summing their energy densities, such that: $T^4_{gas,total} = T^4_{viscous} + T^4_{radmc}$. The details of these calculations are outlined in \S \ref{sec:resTrap} and appendix \ref{sec:app02}.

\subsection{ Evolution of HL Tau's Disk }\label{sec:HLTau}

\begin{table*}
\centering
\caption{Model parameters between the four disk models. Also shown is the gas surface density and temperature at 10$^{5}$ yr and 1 AU, as well as the global mass accretion rate through the disk at 10$^5$ yr.}
\begin{tabular}{|c|c|c|c|c|c|}
\hline
Name & M$_{disk,0}$  & L$_{UV}$ & $\Sigma_0(t=10^5{\rm yr},r=1{\rm AU})$ & T$_0(t=10^5{\rm yr},r=1{\rm AU})$ & $\dot{\rm M}(t=10^5 {\rm yr})$ \\\hline
& M$_\odot$ & L$_{UV,TW~Hya}$ & g/cm$^2$ & K & M$_{\rm sol}$/yr \\\hline
low-mass (LM) & 0.25 & 100 & 1040 & 628 & $2.4\times 10^{-8}$ \\\hline
mid-mass (MM) & 0.44 & 100 & 1565 & 847 & $4.9\times 10^{-8}$ \\\hline
high-mass (HM) & 0.63 & 100 & 2135 & 1024 & $8.1\times 10^{-8}$ \\\hline
low-flux (LF) & 0.63 & 1 & 2135 & 1024 & $8.1\times 10^{-8}$ \\\hline
\end{tabular}
\label{tab:p401}
\end{table*}

For our purpose of testing the possibility of planet formation in  HL Tau, we wish to use a disk model whose properties resemble those of the HL Tau disk. Generally the disk parameter that can most easily be tuned to match observed properties is the initial mass of the disk. Unfortunately since most Class-II systems are up to a few Myr old, this is impossible to constrain observationally for any individual system. Instead we pick disk models with initial masses that evolve to have the same mass as is observed in the HL Tau disk at an age of $\sim 1$ Myr. We used the stellar parameters outlined by \cite{WhiteHill2004}, and a total dust mass of (1-3)$\times 10^{-3}$ M$_\odot$ \citep{CarrGonz2016} to estimate a total gas mass of 0.1-0.3 M$_\odot$ assuming a global gas-to-dust ratio of 100. Assuming that the disk is 1 Myr old then its initial gas mass must have been 0.25-0.63 M$_\odot$ according to the evolution of our gas disk model. Thus, to account for the range of disk masses, we run three models with initial masses of 0.25 M$_\odot$, 0.44 M$_\odot$, and 0.63 M$_\odot$, that we define as the `low-mass', `mid-mass', and `high-mass' models respectively.  These initial disk masses agree with the results of disk formation in star formation simulations carried out by \cite{Bate2018}

Each of these initial masses are larger than any of our previously tested models. And hence, we checked the initial gravitational stability of each of these models. Based on the standard Toomre Q stability analysis we found that initially the low, mid, and high mass models are gravitationally stable within radii of 38, 32, 26 AU respectively. Outside of these radii the disks are gravitationally unstable and will produce mass accretion rates that are larger than are given by viscous evolution alone. These higher accretion rates will raise the temperature of gas which will shift the location of the ice lines outwards from where our model would have placed them. These stability regions move outward as the disk ages, and at 1 Myr the low and mid-mass models have become fully stable, while the high-mass disk model is stable within 46 AU. We expect that our disk model is a good representation of the temperature and surface density profiles of the gas and dust in the gravitationally stable regions of the disk, and hence are useful to study planet formation in the HL Tau system, as well as trapping at planet traps. One possible exception is the CO ice line which resides at large radii ($r > 60$ AU) for all disk models.

The reported UV flux of HL Tau \citep{Bitner2008} is approximately 100$\times$ the observed flux from TW Hya \citep{Yang2012}. So we therefore scale the UV spectrum by a factor of 100 in our radiative transfer scheme. To facilitate comparison to our past work, where we assumed the total UV flux was the same as TW Hya in all models, we ran a second high-mass model where we do not scale the UV spectrum. We summarize the varied parameters in Table \ref{tab:p401}.

\section{ Results: The Positions of Ice Lines and Other Traps in HL Tau} \label{sec:gapsandtraps}

We compute the evolution of the gas and dust surface density and temperature throughout the lifetime of the disk. These evolving profiles change the location of the ice lines as the disk cools and the surface density of material drops, particularly inward of the heat transition. 

\begin{figure*}
\centering
\subfigure[low-mass model]{
\includegraphics[width=0.48\textwidth]{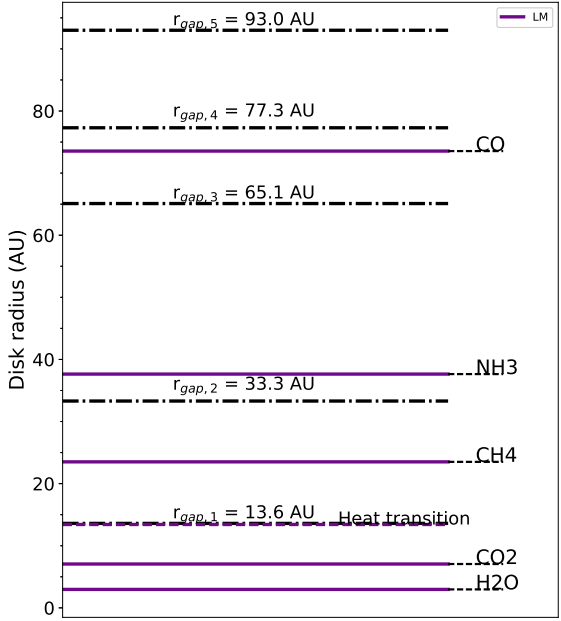}}
\subfigure[mid-mass model]{
\includegraphics[width=0.48\textwidth]{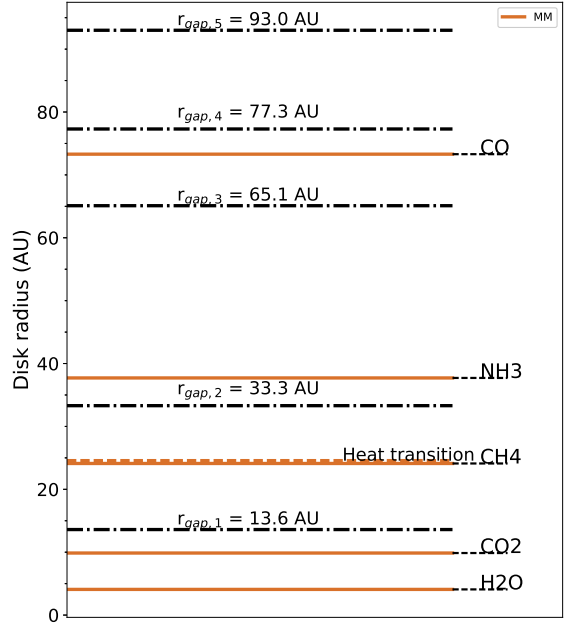}}
\subfigure[high-mass model]{
\includegraphics[width=0.48\textwidth]{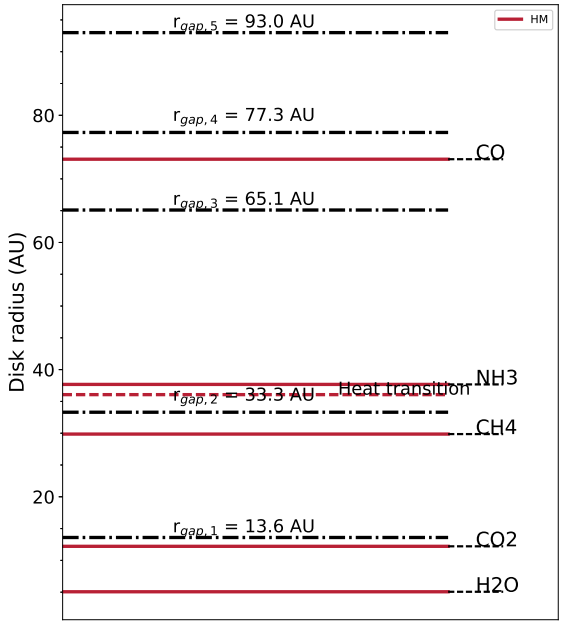}}
\subfigure[low-flux model]{
\includegraphics[width=0.48\textwidth]{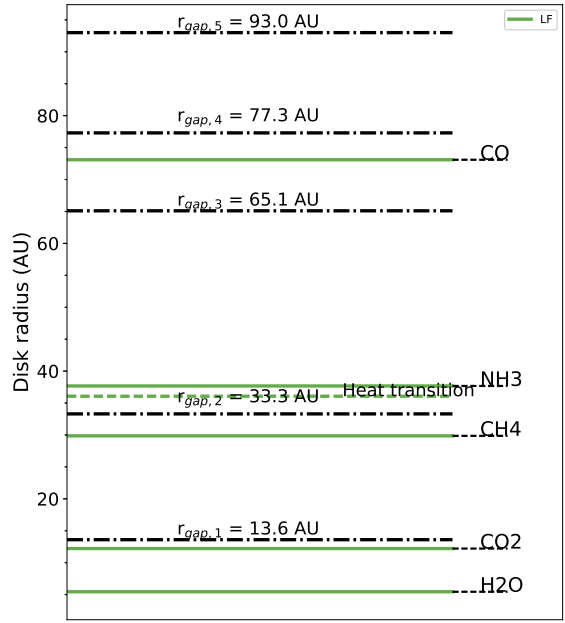}}
\caption{ Radial locations of the condensation fronts of abundant volatiles (coloured solid) and heat transition (colour dashed) for each of our models at 1 Myr, along with the location of the dust gaps (dotted black) in HL Tau.}
\label{fig:p4:results402x}
\end{figure*}

\begin{figure*}
\centering
\subfigure[low-mass model]{
\includegraphics[width=0.48\textwidth]{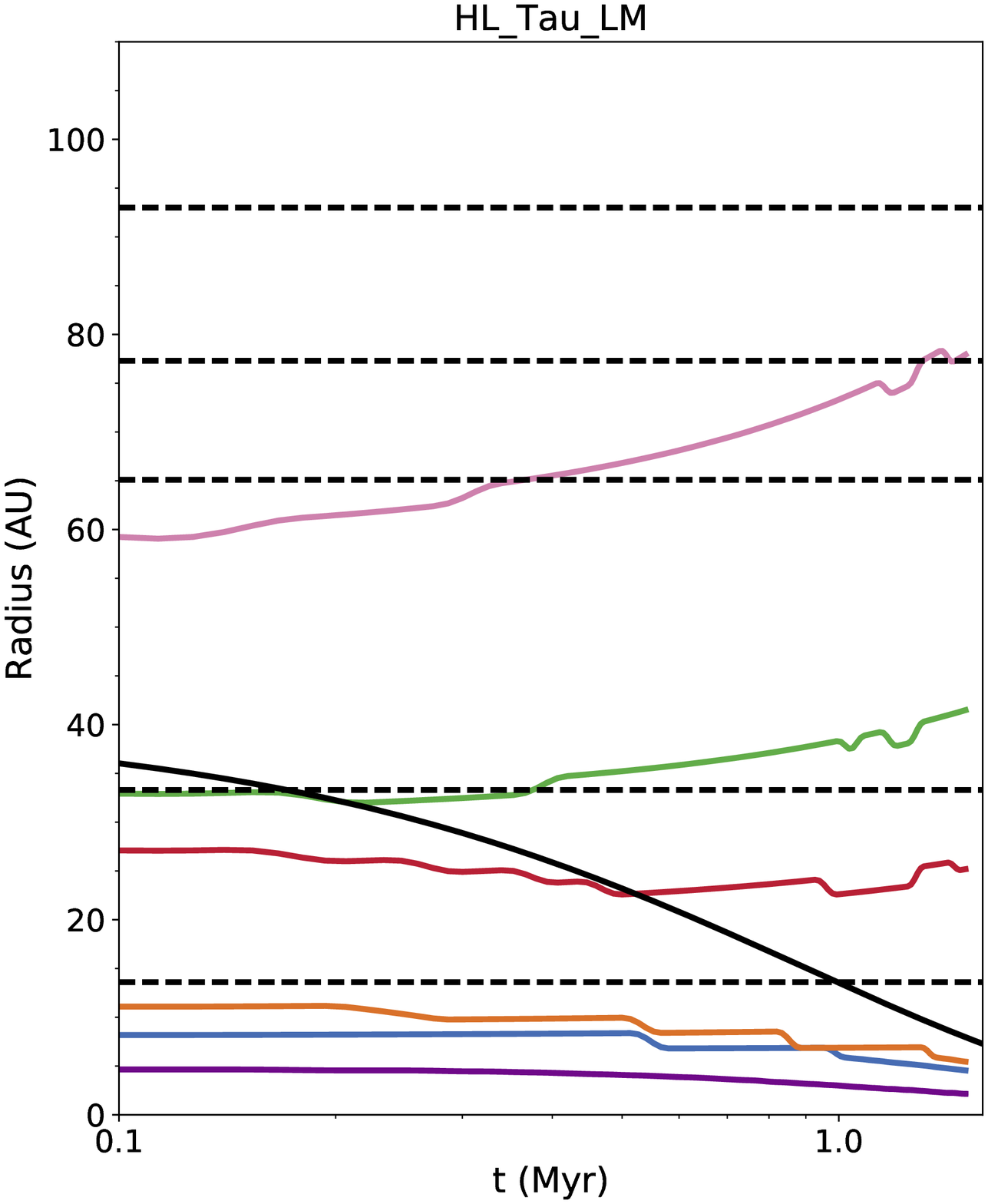}}
\subfigure[mid-mass model]{
\includegraphics[width=0.48\textwidth]{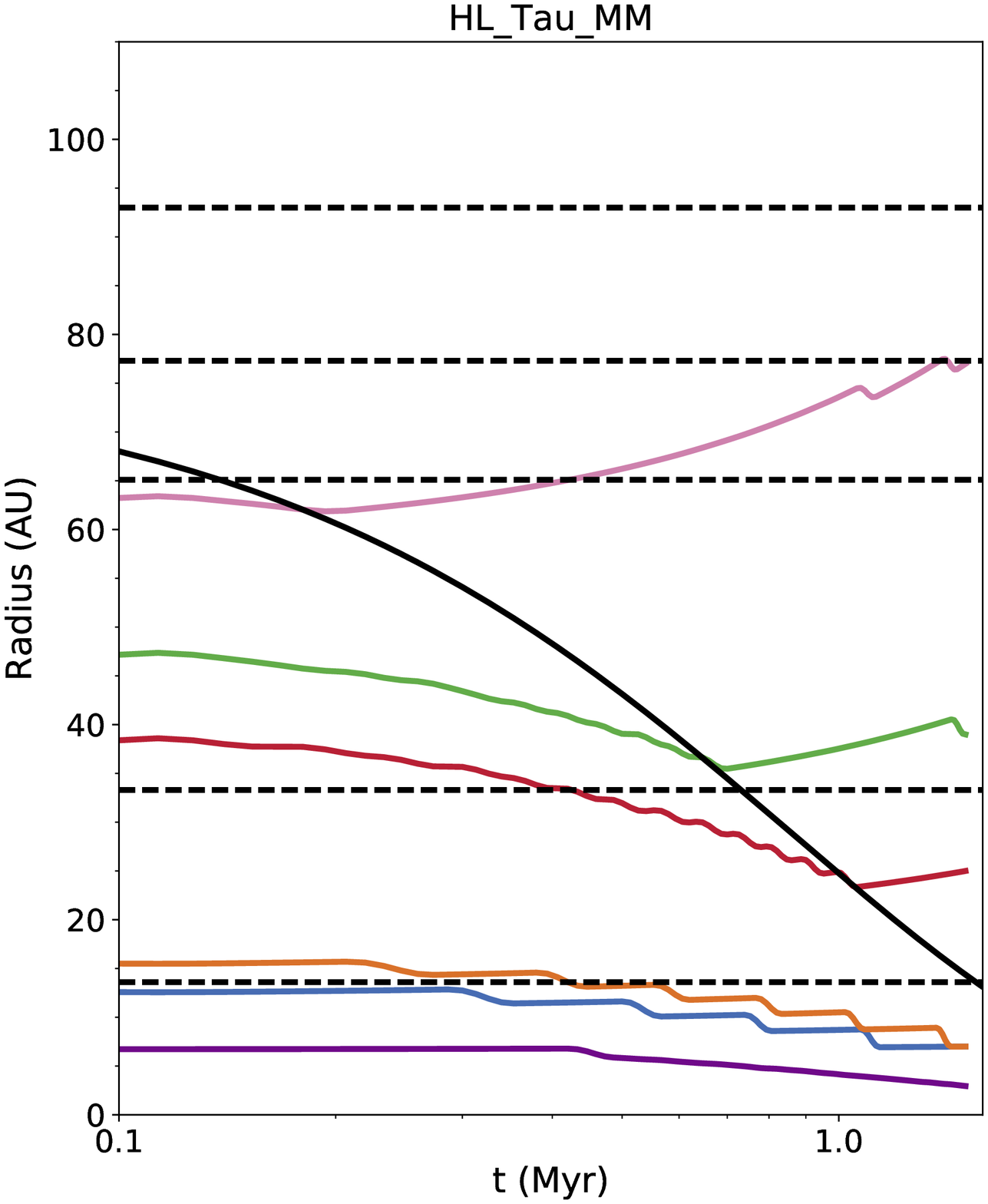}}
\subfigure[high-mass model]{
\includegraphics[width=0.48\textwidth]{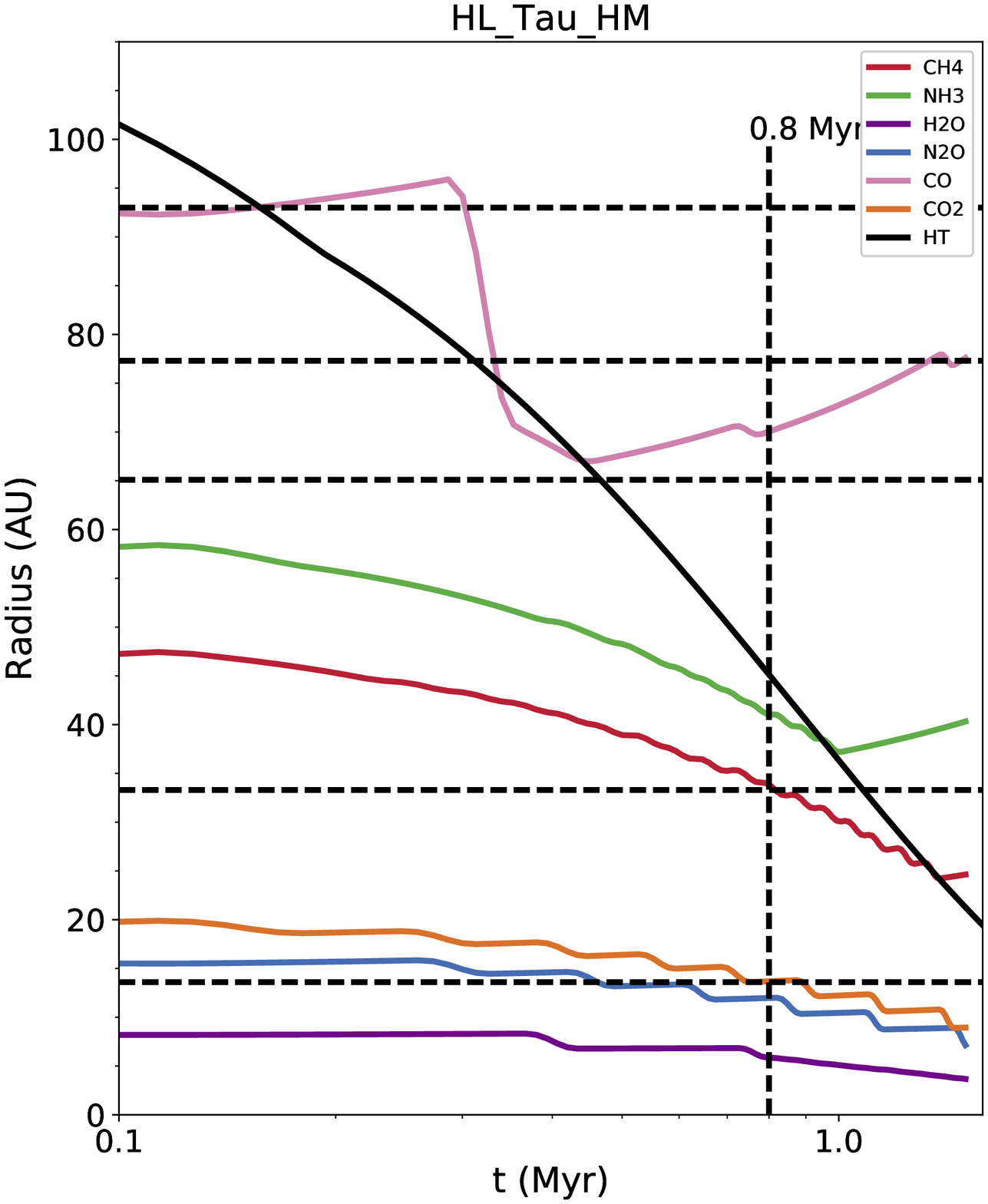}}
\subfigure[low-flux model]{
\includegraphics[width=0.48\textwidth]{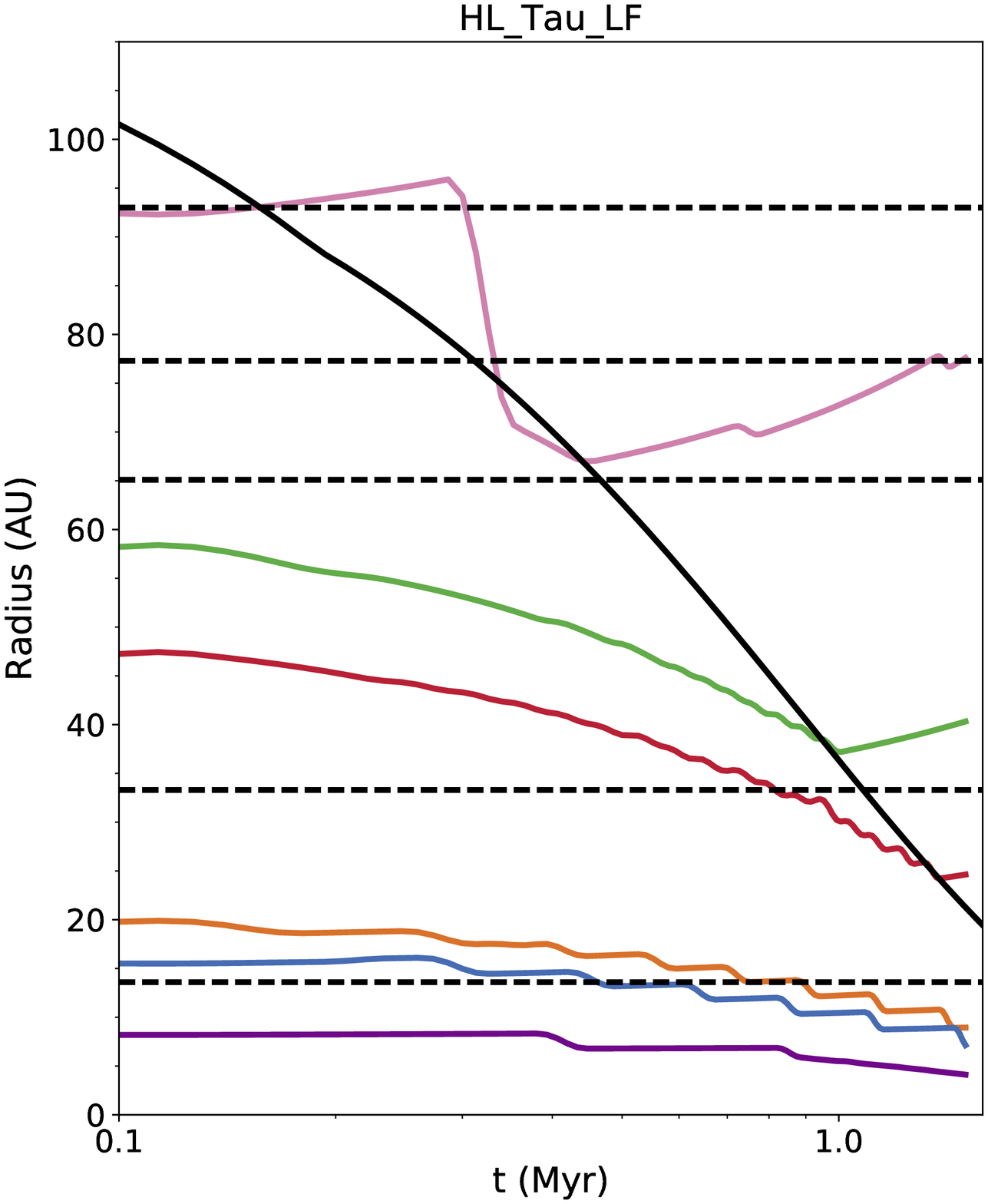}}
\caption{ Temporal evolution of the ice line location in each of the presented disk models. We find that in the high-mass model there is a coincidence between the location of the CO$_2$, CH$_4$ ice lines with the inner two gaps at a disk age of $\sim 0.8$ Myr (marked by a vertical dashed line). A similar coincidence is found in the mid-mass model at an age of $\sim 0.4$ Myr. The evolution of the heat transition (HT) is shown in black.}
\label{fig:p4:results402xx}
\end{figure*}

In Figure \ref{fig:p4:results402x} we compare the radii of the condensation fronts for abundant volatiles in our chemical model and heat transition radius at 1 Myr in each of our disk models to the locations of the dust gaps in HL Tau as reported by \cite{Tamayo15}. 

Generally we find that the CO$_2$ is located closest to the inner gap, with CH$_4$ and NH$_3$ close to the second gap. In the high-mass and low-flux models the CO$_2$ ice line is more coincident with the inner gap than the lower mass models. This difference is due to the higher temperatures in the high-mass and low-flux models, attributed to a higher mass accretion rate and more efficient heating from viscosity. This difference in accretion rates also results in the shifted location of the heat transition between each of the models. Finally in the region of the disk that is primarily heated by direct irradiation (outward of the heat transition) there is no difference in the location of the CO and NH$_3$ ice lines. Because the temperature structure is independent of the amount of material outward of the heat transition when the disk is heated by direct irradiation.

The freeze out of pure CO occurs at the same radius ($\sim 73$ AU) in each of our models, and is located near the fourth gap.  Another abundant volatile, N$_2$ never produces a `traditional' ice line in our model - where the frozen phase becomes more abundant than the gas phase. However its ice phase reaches a maximum at $\sim 90$ AU in each of the models. Because its sublimation temperature is generally slightly lower than CO, its clear that photo-chemical effects like photo-desorption are suppressing the freeze out of N$_2$, pushing its ice line far from the host star. 

\ignore{Based on the surface brightness profile that is reported by \cite{Zhang15} and the fact that the three outer gaps are not near any other ice line in our model, it is unlikely that they are caused by a single planet located at each gap. Recent numerical work has demonstrated that a super-earth planet embedded in a disk similar to HL Tau can lead to the presence of multiple gaps \citep{Dong2017}. This could mean that a planet trapped at the CO ice line could be responsible of at least gaps 3 and 4.}

The coincidence of the dust gaps and condensation fronts has been previously discussed in \cite{Zhang15} who used the temperature profile $665 (r/{\rm AU})^{-0.6}$ K to determine the radial location of the ice lines. This profile was derived by a 2D radiative transfer model designed to model the observed SED of the HL Tau system \citep{Men1999}. At 1 Myr the temperature profile in our high-mass disk model is $630 (r/{\rm AU})^{-0.9}$ K when $r< r_t$ and $24 (r/40{\rm AU})^{-0.42}$ K when $r > r_t$. We see that in our model the gas is almost always cooler than the model used by \cite{Zhang15}, other than very near ($r<0.8$ AU) to the host star. This results in our disagreement of which ice lines are more coincident with the gap locations (see below).

\cite{Zhang15} report that the location of the inner two gaps correspond to the water and hydrated NH$_3$ ice lines respectively. They report the start of a large outer gap at the CO$_2$ ice line and is centered on the sublimation temperature of a combination of CO and N$_2$ on a water ice mantle. This large gap spans the radii between the third and fourth gap reported by \cite{Tamayo15} and used here. The fifth gap is not represented in \cite{Zhang15}. 

An important distinction between the results of \cite{Zhang15} and our own work is that they considered the freeze out of hydrated species of volatiles, as well as the freeze out of volatiles on already established icy mantles - resulting in higher binding energies, and hence higher sublimation temperatures than pure volatile freeze out. This extra layer of complexity is generally not included in astrochemical simulations, and could cause small radial shifts in the location of the ice line. Since we model the temporal evolution of the disk, any small shift in radius becomes equivalent to uncertainty in the system age.

We can search for a particular epoch that may show better correspondence between ice lines and the gaps found in Figure \ref{fig:p4:results402x}. In Figure \ref{fig:p4:results402xx}, we show the time evolution of the ice line location for the most abundant volatiles and compare their location to the dust gaps in the HL Tau disk. Note that in the high-mass and low-flux models the CO ice line does not initially exist, because the gas starts too hot for CO to freeze out. In those models, for plotting convenience we simply set the CO ice line to equal the disk edge. As a result, the curve initially evolves outward as the disk spreads, then quickly moves inward as the disk cools and an ice line appears. Similarly, we include the estimated location of the heat transition at all times, even if (in the case of the `high-mass' model) it is not initially within the disk radius.

Note an outward motion for ice lines that cross the heat transition (black curve). Once an ice line is outward of the heat transition it exists in a region of the disk that has no evolution in its temperature profile - because heating is dominated by (an assumed) constant radiation field. As the surface density of the disk drops, photodesorption begins to impact the desorption rates of ice species, pushing the ice line outward.

By following their temporal evolution, we look for an epoch when the ice lines best agree with the location of gaps in the disk, as an additional method of estimating the system age. The ice line evolution is largely determined by the thermal evolution of the gas disk, and hence is given by the analytic disk model that we use. As a result, the high-mass and low-flux models have identical ice line evolution since their disk started with the same initial mass. In what follows we will exclusively discuss results as they apply to the high-mass model while implying that all conclusions can similarly be applied to the low-flux model. 

Both the high-mass and mid-mass models show an instance when multiple ice lines coincide with the gaps. These epochs occur at ages of $\sim 0.8$ Myr and $\sim 0.4$ Myr respectively. Because the latter age is unreasonably small for the HL Tau system, and since the former age lies closest to the assumed age of the HL Tau disk ($\sim 1$ Myr) we will favour the disk parameters of the high-mass model in our analysis of planet trapping at ice lines.  

\section{ Results: Planet Trapping at Ice Lines }\label{sec:resTrap}

The process of planet trapping at ice lines is driven by an opacity transition which locally changes the power law index of the temperature and gas density profiles. To sufficiently change the opacity across an ice line, one would expect that the volatile must be abundant. Water is an excellent candidate for trapping because of its abundance and because  its low average density has a strong effect on the total opacity of the solids when it freezes out (see \cite{Miyake1993} and Appendix \ref{sec:app01}). Here we demonstrate planet trapping due to the opacity transition at the water ice line by modelling the freezes out of water as a function of radius across the ice line, and computing the resulting temperature profile for a viscously heated disk. In what follows we will use the stellar and disk parameters used in the `high-mass' disk model at the epoch of highest coincidence between ice line and gap locations ($t \sim 0.8$ Myr), unless otherwise specified. This model will act as our fiducial model, and provide a base from which we build our modified disk model (see below).

To assess the impact of the water ice line on the total torques on a planet we have added a much deeper treatment of opacity effects to our earlier work. It is based on the analytic solutions of \cite{Cham09}, and used in our previous work \citep{Crid16a}. This new enhanced (denoted `modified') model takes the radial distribution of water ice from our previous astrochemical results (see Figure \ref{fig:p4:new01x}) as an input to compute the change in dust opacity (and subsequent temperature profile) across the ice line using details of grain properties. In particular our opacity model depends on the mass abundance of ice that has accumulated on the grain - a property that depends on disk radius across the ice line. The details of this calculation are presented in Appendix \ref{sec:app01}. Above a temperature of 1380 K, where dust is expected to sublimate, we return the prescription of the temperature depended opacity as discussed in \cite{Cham09}.

Our method of deriving the radial dependence of the icy dust opacity differs from other theoretical efforts like \cite{BellLin1994} or \cite{St98} (among others). In these works, the opacity is described by a power-law of temperature and density over different temperature ranges, depending on the dominant physical effect generating the opacity \citep{BellLin1994}. These power-laws are fit or compared to tabulated values\footnote{to \cite{Alexander1989} in the case of \cite{BellLin1994}} and the transition between temperature regions are either smoothed \citep{BellLin1994,St98} or not \citep{Baillie2016}.
	
In this work we combine the tabulated opacities of bare silicates and pure water ice into a single Planck mean opacity using the method described by \cite{Miyake1993} (also see Appendix \ref{sec:app01}). This method computes the effective complex spectral index for a mixture of ice and dust over a range of wavelengths. The relative abundance of ice and dust is set by our astrochemical disk model and depends on radius. These indices are then used to compute the total opacity using the internal opacity calculator of RADMC3D (for more details see Appendix \ref{sec:app01}). In this way we can control the impact of a radially changing abundance of water ice without having to assume any power-law functional form for the temperature dependence in the opacity.

\begin{figure}
\includegraphics[width=0.5\textwidth]{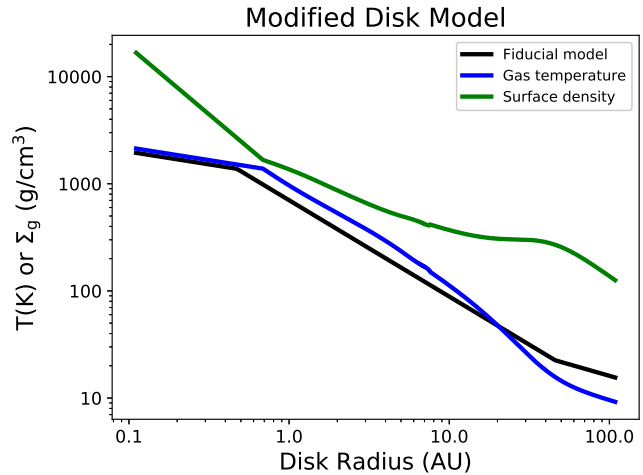}
\caption{The gas temperature and surface density profiles for the modified disk model used to illustrate planet trapping. The initial temperature profile was taken from the `high-mass' model at t = 0.8 Myr and denoted as `Fiducial model'.}
\label{fig:p4:new02}
\end{figure}

In Figure \ref{fig:p4:new02} we show the resulting midplane temperature and surface density profiles for the modified disk model. The ice line is within the viscously heated region of the disk, and the resulting drop in opacity allows for more efficient cooling. Therefore, the midplane temperature profile becomes slightly shallower within the ice line before returning to a slightly steeper power-law outward of the ice line. Because the accretion rate is constant in steady state disk theory, this restricts the surface density profile (see Appendix \ref{sec:app01}). One might worry that these modifications might effect the thermal stability of the disk. We tested the Rayleigh stability of the disk (e.g. \cite{YangMenou2010})\footnote{The Rayleigh stability requires $V_k^2 + \frac{1}{r}\frac{\partial}{\partial r}\left(\frac{r^3}{\rho}\frac{\partial P}{\partial r}\right) > 0$, where $P = \rho c_s^2/\gamma$ is the gas pressure, $\rho$ is the gas volume density, and $V_k^2 = GM_*/r$ is the square of the Kepler speed} and find that our modified disk model does not harm the thermal stability of the disk.

We note that this modified disk model has a similar radial dependence inward of the ice line for both the temperature and density as in our fiducial disk model. The temperature profile steepens outside of the ice line as the temperature of the gas and dust is reduced, then eventually flattens as heating from direct irradiation becomes dominant. The disk is truncated at its maximum radius ($\sim 112$ AU at t = 0.8 Myr) in line with our previous work which does not include an exponential taper. This maximum radius is determined by the viscous spreading associated with the inward mass flux through the disk (see \cite{Cham09} for details).

The dust opacity is higher inward of the ice line so the cooling is less efficient resulting in a higher temperature at low radii. In principle these higher temperatures will shift the location of the water ice line outward in the disk. However this shift is small and does not effect the total torque on a planet.  This model includes the change of the temperature radial profile at the heat transition, where gas heating becomes dominated by radiative processes rather than viscosity (outside of R$\sim 30$ AU). This is discussed in more detail below.

\begin{figure}
\begin{overpic}[width=0.5\textwidth]{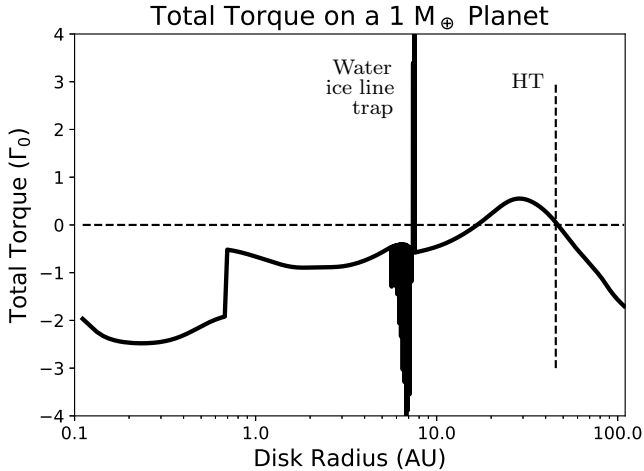}
\put(78,60){HT}
\put(50,62){ Water }
\put(49,59){ ice line }
\put(53,56){ trap }
\end{overpic}
\caption{Total torque acting on a 1 M$_\oplus$ planet in our modified disk model normalized by $\Gamma_0$. A strong positive total torque appears at the ice line, signifying trapping, while a strong transition appears near the heat transition (HT, vertical dashed line).}
\label{fig:p4:new03}
\end{figure}

In Figure \ref{fig:p4:new03} we show the total torque acting on a 1 M$_\oplus$ planet in our modified disk model. The total torque includes both the Lindblad torques as well as the co-rotation torques, and were computed using the same method as \cite{Cole14}, based on the work of \cite{Paard2011} (see Appendix \ref{sec:app03} for a brief summary). The total torque is normalized by:\begin{align}
\Gamma_0 = (q/h)^2\Sigma_g r_p^4 \Omega_p^2
\end{align}
where $q = M_p/M_*$ and $h = H/r_p$. The strong positive torque at the ice line signifies the trapping that would occur there, and is determined by the change in the temperature profile across the ice line. We note the fiducial location of another assumed planet trap, the heat transition (vertical dashed line). 

Also noted in Figure \ref{fig:p4:new03} is the location of the heat transition (HT) - where the disk heating transitions from being dominated by viscosity to being dominated by the direct irradiation of the host star. At the heat transition there is also evidence for planet trapping, where the total torque transitions from positive to negative. This confirms our previous assertion (see for ex. \cite{Crid16a}) that the heat transition should act as a planet trap.

It is noteworthy that there is another null point in the total torque between the ice line and heat transition traps. It can be shown that this null point is actually unstable, and rather represents a position from which a planet would be repelled rather than trapped.

\subsection{ Varying Water Ice Abundance }

The change in dust opacity across the ice line is determined by the abundance of water ice that is frozen out. Here we test to what degree the change in gas temperature profile is controlled by the maximum water ice abundance that freezes out, and how this is reflected in the net torque on a planet.

\begin{figure}
\begin{overpic}[width=0.5\textwidth]{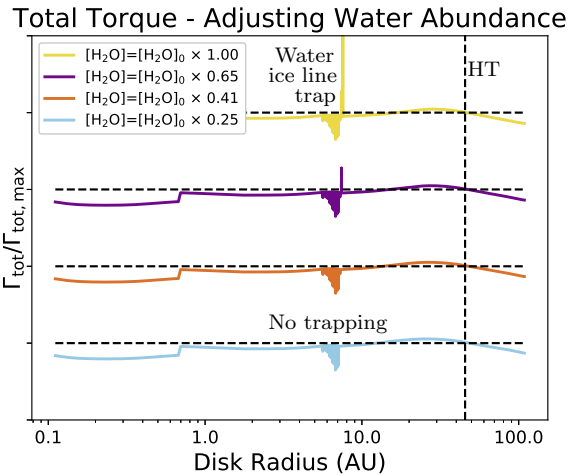}
\put(78,60){HT}
\put(48,62){ Water }
\put(47,59){ ice line }
\put(51,56){ trap }
\put(47,22){ No trapping }
\end{overpic}
\caption{Variation in the maximum water ice abundance relative to the fiducial water abundance. After a drop of about half of the ice abundance the planet trapping signature disappears.}
\label{fig:varwater01}
\end{figure}

In Figure \ref{fig:varwater01} we show a set of total torques where we vary the maximum allowed water ice abundance on the grains. Each curve is shifted, and is normalized to the maximum torque of the fiducial model (yellow line). Each horizontal dashed line shows the zero-torque point for each ice abundance test, and the vertical dashed line shows the location of the heat transition. As the maximum abundance of ice is reduced, the height of the positive torque at the water ice line shrinks, eventually shrinking below zero when the water abundance is 40$\%$ of fiducial value. Hence the planet trapping switches off when the water abundance drops below roughly half of the maximum water abundance from the fiducial model. 

This result differs from those of \cite{Bitsch2016}, who found no such reduction of planet trapping for a drop in abundance of an order of magnitude. However here we begin with a lower ice-to-silicate mass ratio, implying that the ice-to-silicate ratio must be below at least about 10\% before seeing such an effect.

This result implies that volatiles that are less abundant than about half of the water abundance from the fiducial model will not produce planet traps at their ice line. Additionally our result implies that protoplanetary disks that either inherited low abundances of water, had their water destroyed early in their evolution (ie. the reset model of \cite{Eistrup2016}), or during the collapse stage \citep{Drozd16} could have less efficient planet trapping at their water ice line.

\begin{figure}
\begin{overpic}[width=0.5\textwidth]{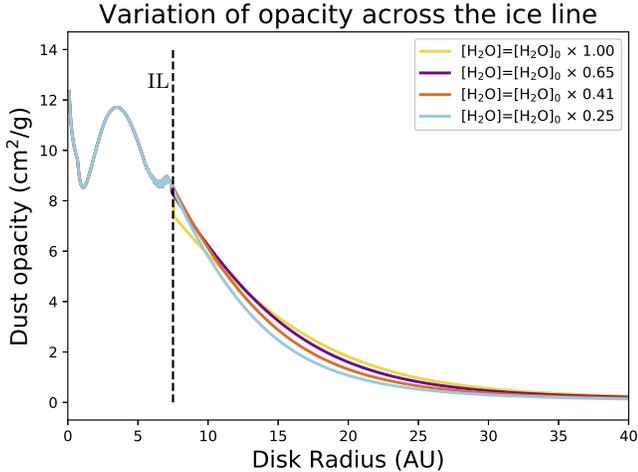}
\put(23,60){ IL }
\end{overpic}
\caption{ Radial dependence of the dust opacity for the four cases with varying maximum water abundances. The opacity drop at the water ice line is less severe as the maximum water abundance is reduced, hence the ability of the ice line to trap planets is reduced. We note the location of the water ice line (where the water ice abundance exceeds half of its maximum abundance in the fiducial model) with a dashed line.}
\label{fig:varwater02}
\end{figure}

In Figure \ref{fig:varwater02} we show the radial dependence of the dust opacity over the disk (up to a radius of 40 AU) as computed in Appendix \ref{sec:app01}. We see why in Figure \ref{fig:varwater01} the trapping disappears as the water abundance drops - because the reduction of opacity is less severe across the ice line as the water ice abundance is reduced. Hence for ice species with spectral features similar to water and abundances half that of water (in our fiducial model), we would not expect a sufficiently high opacity drop to lead to planet trapping. 

Because of the multiplicity of ice lines in the HL Tau system, an interesting question arises: can the ice lines of other volatiles also act as planet traps? Below we test both CO$_2$, which is much less abundant than water in our model; and CO which is as abundant but resides in a part of the disk that is heated through irradiation (rather than viscosity).

\subsection{Planet Trapping at the CO$_2$ Ice Line?}

\begin{figure}
\centering
\begin{overpic}[width=0.5\textwidth]{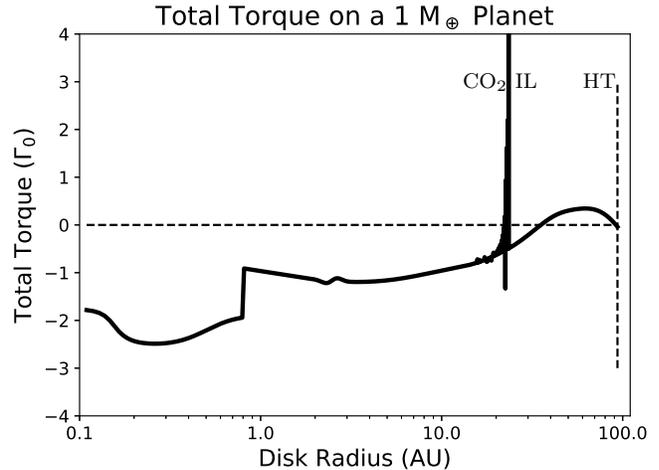}
\put(88,60){HT}
\put(70,60){CO$_2$ IL}
\end{overpic}
\caption{Total torque on a 1 M$_\oplus$ massed planet for the case where dust opacity is determined by the abundance of CO$_2$ ice on water ice-covered dust grains. We additionally note the CO$_2$ ice line (CO$_2$ IL) and the heat transition (HT). Here we show the results for the high-mass model at a time of 0.1 Myr rather than our fiducial time of 0.8 Myr, hence the heat transition is farther out than in previous figures. }
\label{fig:CO2iceline}
\end{figure}

In Figure \ref{fig:CO2iceline} we show the total torques around the CO$_2$ ice line (CO$_2$ IL) for a 1 M$_\oplus$ mass planet. Similar to the case of the water ice line, the opacity transition at the CO$_2$ ice line causes a strong positive torque. We similarly see planet trapping near the heat transition (outer dashed line), however it is shifted slightly with respect to the true location of the heat transition. This shift comes from the details of the modified disk model around the CO$_2$ ice line, which predicts a slightly warmer disk than was computed for the disk model which varied water ice abundance. As a result the location of the heat transition in this disk model is shifted outward slightly.

In our model CO$_2$ in under-abundant compared to H$_2$O by a factor of about 400, so at first glance, it is surprising given above that a planet can be trapped at its ice line. Further analysis, however, shows that in the modified disk model the temperature profile of the gas is steepening near the CO$_2$ ice line and hence even a relatively small (compared to the effect of water) change to the dust opacity from the freeze out of CO$_2$ flattens the temperature profile enough to lead to planet trapping. In other chemical models \citep{Drozd14,Walsh2014,Eistrup2016} CO$_2$ is more abundant, because they include dust grain surface reactions which lead to its efficient formation, or because CO$_2$ is inherited by the disk and not efficiently destroyed. In these models the opacity transition would be stronger, further supporting the role of the CO$_2$ ice line as a planet trap.

\subsection{ Planet Trapping at the CO Ice Line?}\label{sec:COtrap}
The only other molecule that has similar abundances to H$_2$O in our chemical model is CO. However since it resides in a region of the disk that is heated primarily by direct irradiation by the host star, we require further analysis than is presented in this section to determine if it can indeed trap planets.

The above analysis assumes that the ice line resides in a region of the disk where heating is dominated by viscous dissipation. For the case of the CO ice line however, which lies at radii $\gtrsim 30$ AU, the disk gas will be primarily heated through the direct irradiation of the host star. Because of this we search for the effect of trapping by computing the temperature profile numerically with a Monte Carlo radiative transfer scheme. 

In particular we search for a similar form of the temperature profile arising from the change in dust opacity across the CO ice line as was seen in the viscously heated part of the disk, leading to planet trapping. To test the validity of trapping planets at the CO ice line we similarly compute the dust opacity as a function of radius, starting from a mixture of silicate-water ice and adding CO ice with mass abundances given by the results of our previous astrochemical simulations (see \cite{Crid17} for an example). We follow the same method as before to compute the Planck mean opacity of the dust-ice mixture as a function of radius across the CO ice line (see Appendix \ref{sec:app01}).

To assess the impact of changing the opacity across the CO ice line in a radiatively heated disk we compute the dust temperature of the dust using the radiative transfer code {\it RADMC3D} \citep{RADMC}. The underlying density of the dust and gas are computed assuming solely viscous heating, and a constant gas-to-dust ratio. Hence the temperature computed numerically represents the excess heat available to the dust from stellar radiation, above the effect of viscous heating. For this calculation we simulated 20 dust populations which each have a varying abundance (relative to the number of hydrogen atoms) of CO ice between zero and $2.4\times 10^{-3}$. Each dust grain is the same size (0.1 $\mu$m) and we used 10 million photon packets with {\it RADMC3D}.

\ignore{
\begin{figure}
\centering
\begin{overpic}[width=0.5\textwidth]{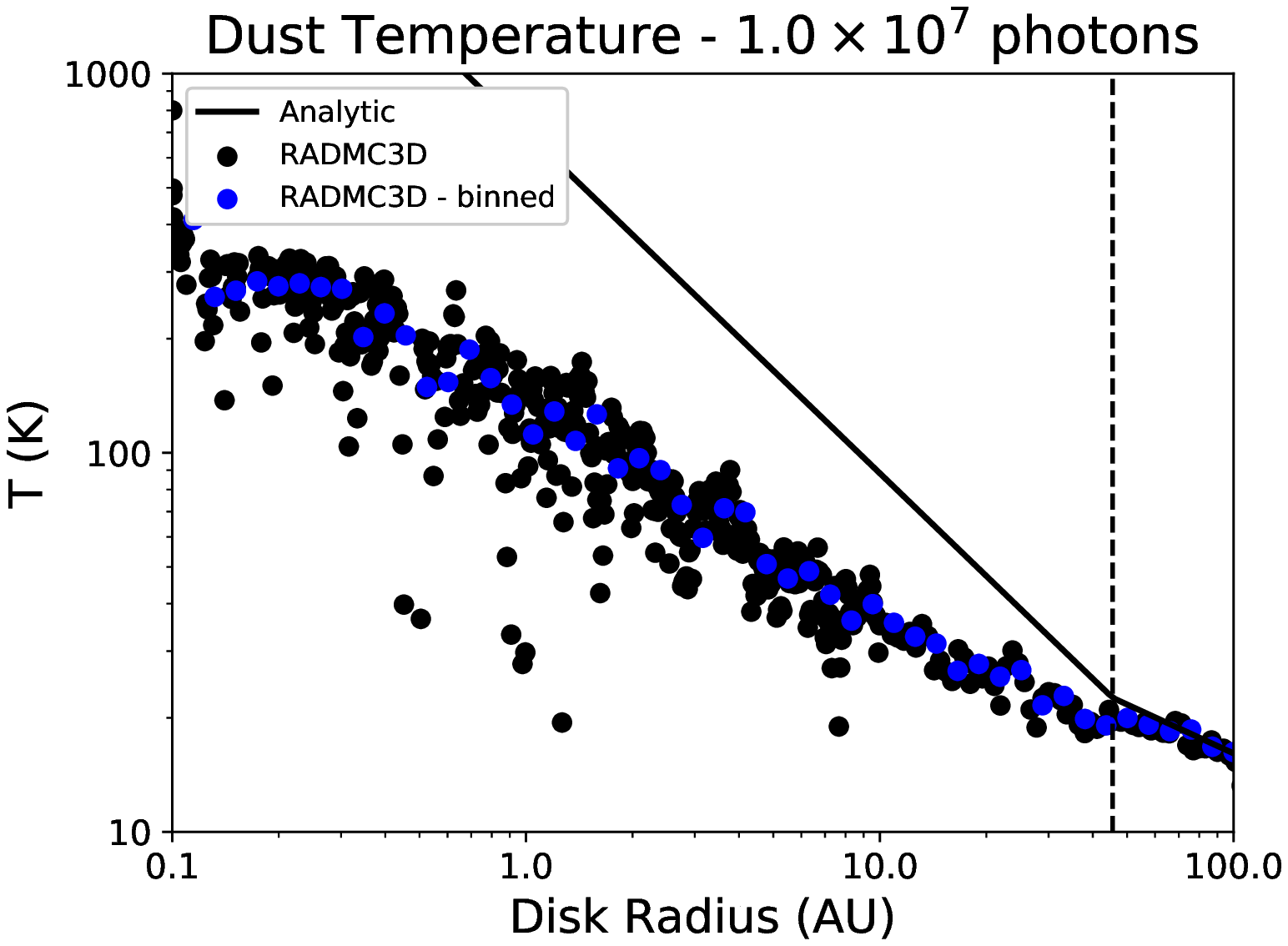}
\put(78,60){HT}
\end{overpic}
\caption{Solid line: midplane dust temperature based on our analytic model. Inward of the heat transition (dashed line) the disk is heated by viscous heating, while beyond the heat transition the midplane is heated by radiative heating. Points: midplane dust temperature from radiative heating alone computed by {\it RADMC3D} (black) as well as the binned data used in torque calculations (blue).}
\label{fig:p4:new06}
\end{figure}
}

\begin{figure*}
\centering
\subfigure[Single dust grains size (grain radius $= 0.1 \mu$m) \label{fig:p4:new06}]{
\begin{overpic}[width=0.5\textwidth]{radmc_temp_9_5.899.eps}
\put(78,60){HT}
\end{overpic}
}%
\subfigure[Full grain size distribution (see Figure \ref{fig:glbl01}) \label{fig:glbl02}]{
\begin{overpic}[width=0.5\textwidth]{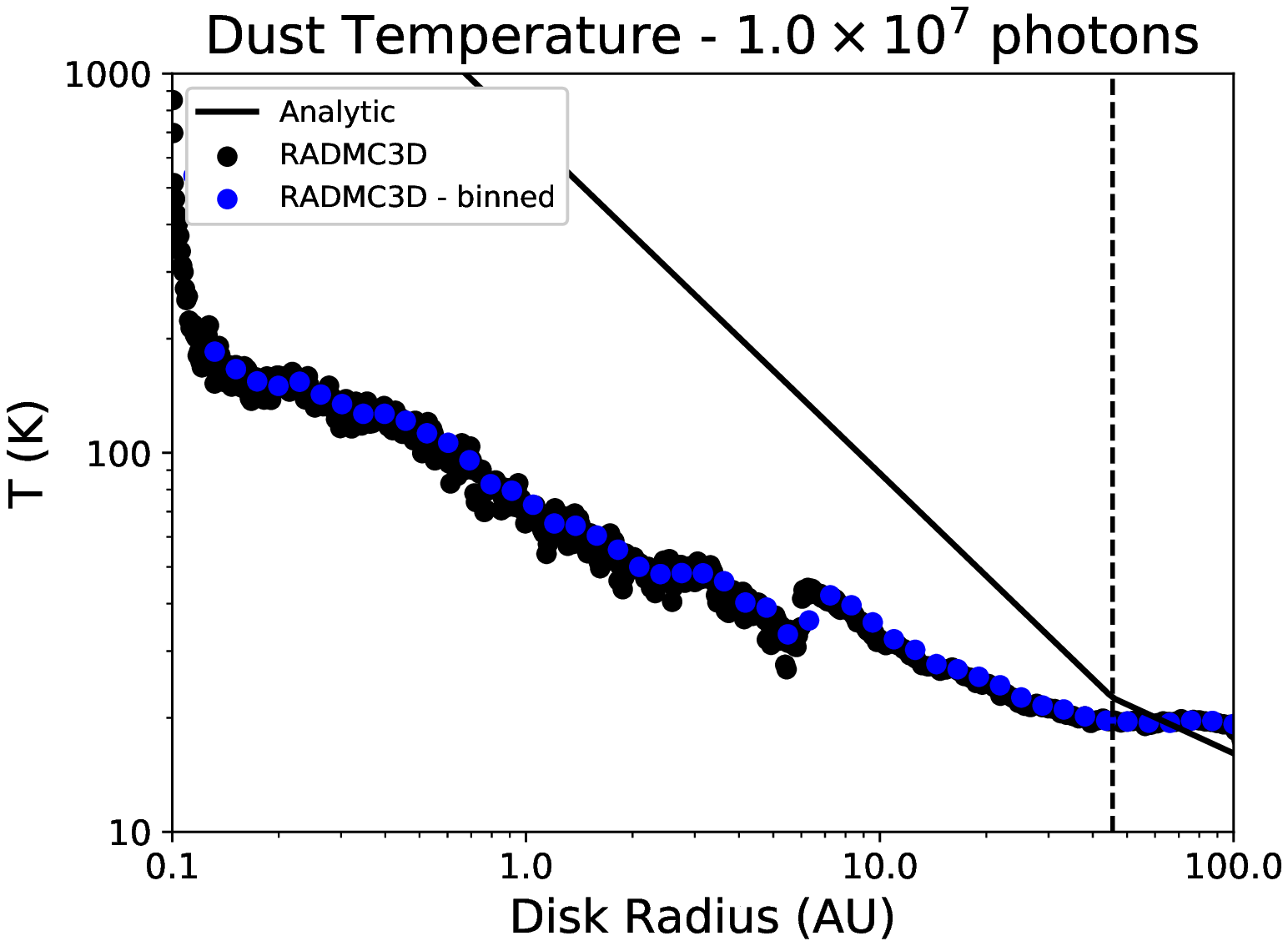}
\put(78,60){HT}
\end{overpic}
}
\caption{Comparison between different dust distribution models, and their resulting dust temperature from \textit{RADMC3D}. Solid line: midplane dust temperature based on our analytic model. Inward of the heat transition (dashed line) the disk is heated by viscous heating, while beyond the heat transition the midplane is heated by radiative heating. Points: midplane dust temperature from radiative heating alone computed by {\it RADMC3D} (black) as well as the binned data used in torque calculations (blue). The left figure is used in \S \ref{sec:COtrap} while the right figure is used in \S \ref{sec:GlobTrap}. The difference in scatter between the two points is related to a net loss of opacity in the right figure when the bulk of the dust mass is allowed to grow from 0.1 $\mu$m (left) to predominately mm-sized grains (right, also see Figure \ref{fig:glbl01}).}
\label{fig:radfields}
\end{figure*}

In Figure \ref{fig:radfields} we compare the midplane dust temperature profiles derived by {\it RADMC3D} (ie. radiative heating, points), and our analytic model (eqs. \ref{eq:p402x} and \ref{eq:p402}, solid). As expected, the dust temperature computed by {\it RADMC3D} recovers the radial dependence of our analytic model outside of the heat transition ($r_t$, dashed line). While within the heat transition we find that the dust temperature is consistently lower than is expected by viscous heating. This conclusion is consistent with the assumption of our analytic model as described by \cite{Cham09}, which asserts that within the heat transition (r$_t$) the dust heating is dominated by the heat released by the viscous evolution of the gas.

We investigate two dust models and their consequences on the heating of the outer disk, and trapping at the CO ice line. The first is that the dust mass is dominated by the smallest (0.1 $\mu$m) grains. Such an assumption overestimates the total dust opacity in the inner ($r < r_t$) region of the disk which causes the scatter observed in Figure \ref{fig:p4:new06}. In principle this scatter would be reduced by the diffusive properties of the radiation field (which is not modelled by {\it RADMC3D}), and is similarly reduced when the numerical data is binned (blue points). For our second case, we compute the full dust size distribution using a Two-pop model (Figure \ref{fig:glbl02}, also see Figure \ref{fig:glbl01}). The bulk of the dust mass is in mm-sized grains, and the total dust opacity throughout the disk is lower. Hence along the midplane the scatter in the dust temperature is lower (see Figure \ref{fig:glbl02}).

In Figure \ref{fig:glbl02} we show the dust temperature derived by {\it RADMC3D} using the dust distributions computed by the Two-population model. Near the dead zone edge we find a short positive gradient in the radiative temperature gradient (points). Likewise, outside of the heat transition the flattened temperature profile is a result of the enhancement of medium and small grains in the outer parts of the disk.

As mentioned above, we compute the temperature radial profile of the disk by combining both viscous and radiative heating profiles. To do this, we first bin the numerical data into 50 radial bins (blue points in Figure \ref{fig:radfields}) spaced evenly in log-space over the range $R \in [0.1,100] {\rm AU}$, and compute the average temperature within each bin (blue points). Next we add the energy densities which result from the viscous and radiative heating ($aT_{vis}^4$ and $aT_{rad}^4$ respectively), such that the final gas temperature is:\begin{align}
T_{tot} = \left(T_{vis}^4 + T_{rad}^4\right)^{1/4},
\label{eq:Tpro01}
\end{align} 
where $T_{vis}$ is given by our analytic model (case $r<r_t$ in equation \ref{eq:p402}), and $T_{rad}$ is given by the binned numerical data). This combination assumes that the dust and gas are in thermal equilibrium, which is a reasonable assumption along the midplane of the disk. Such a combination produces a temperature profile similar to the analytic model plotted in figure \ref{fig:p4:new06}, but smooths the transition between viscous and radiative heating. As before, we compute the associated surface density profile assuming that the global mass accretion rate through the disk is constant in space ($\mdot = 3\pi\nu\Sigma$), and the viscosity is given by the standard $\alpha$-disk model ($\nu = \alpha c_s^2 / \Omega$).

\ignore{
In Figure \ref{fig:p4:new07} we show the total torque on a 1 M$_\oplus$ planet as a function of radius. In computing the torques here we have ignored the effect of the water ice line. We once again find that the heat transition does not produce planet trapping in our disk model. Similarly, we do not find evidence for planet trapping at the CO ice line.}

We find no reversal in the direction of the net torque, and hence there is no trapping at the CO ice line. The lack of trapping comes from the fact that the dust opacity is not greatly reduced (factor of order unity) as CO freezes onto the grains. The opacity transition is therefore insufficient at the CO ice line to produce a strong positive net torque.

We have shown that the inner two gaps lie near the CO$_2$ and CH$_4$ ice lines and that CO$_2$ is sufficiently abundant in our model to lead to the trapping of a 1 M$_\oplus$ planet. However, the CO ice line does not show evidence of planet trapping because the freeze out of CO does not constitute a large enough reduction in dust opacity (factor order unity) to lead to a perturbed temperature profile in the radiatively heated part of the disk. The CH$_4$ ice line is similarly	 located near the heat transition where, along with its low abundance we do not expect it to exhibit planet trapping. The third and fourth gaps surround the location of the CO ice line, however it is unlikely that these features are due to a planet trapped at the CO ice line.

\section{ Results: A Global Picture of Planet Trapping}\label{sec:GlobTrap}

\subsection{ Including the dead zone and heat transition }

With the above analysis we are now in a position to study the trapping details around the other planet traps that have appear in our past work (ie. \cite{Crid17}) as well as in the planet formation model in section \ref{sec:resPlnt4}. The two other planet traps are the dead zone edge and the heat transition, representing a transition in the turbulent $\alpha$ ($\alpha_{turb}$) and primary gas heating source respectively.

Trapping at the dead zone edge has been attributed to a rising midplane temperature profile caused by the efficient settling of grains \citep{HP10}. The increase in solid density blankets the midplane from outgoing radiation, producing a positive temperature gradient. Likewise, the heat transition trap results from a change in the temperature gradient due to radiative heating at the midplane becoming more efficient than viscous heating (ie the second term in equation \ref{eq:Tpro01} becoming larger than the first).

As before, these trapping mechanisms depend on the temperature profile of the disk. However in this section the interplay between the radial distribution of the dust and the radiation field is more important than the radial distribution of the ice. Hence we replace our previous assumption of the dust grain population being dominated by a single grain size for a semi-analytic description of the surface density of grains with different sizes. For simplicity, and since the bulk of the disk is exterior to the water ice line, we keep the dust opacity constant in the disk during the Monte Carlo calculation of the dust temperature. Furthermore, since the water ice line is in the viscously heated part of the disk we do not see a strong contribution to the temperature profile from radiation (for example see Figure \ref{fig:p4:new06}), hence changing the opacity across the ice line will not have a large impact on the dust temperature.\footnote{We assume that the grains are all covered by layer of water ice, with a constant average opacity of $\kappa$ = 3 cm$^2$/g}.

We combine the Two-population dust model outline in \cite{Crid16b} (based on \cite{B12} and \cite{twopoppy}) with {\it RADMC3D} to compute a new temperature profile and the resulting total torques around the dead zone edge and heat transition.

In the Two-population model the coagulation, fragmentation, diffusion, and radial drift of dust is computed numerically using two representative dust populations - resulting in the total dust surface density radial profile. Then the surface density radial profile for a range of dust sizes are reconstructed using semi-analytic expressions which describe the average evolution of the grains as a function of their size. We implemented the evolution and implication of the dead zone in the Two-population model in a similar way as in the Appendix of \cite{Crid16b} (also see Appendix \ref{sec:app02} in this work). 

\begin{figure}
\centering
\begin{overpic}[width=0.5\textwidth]{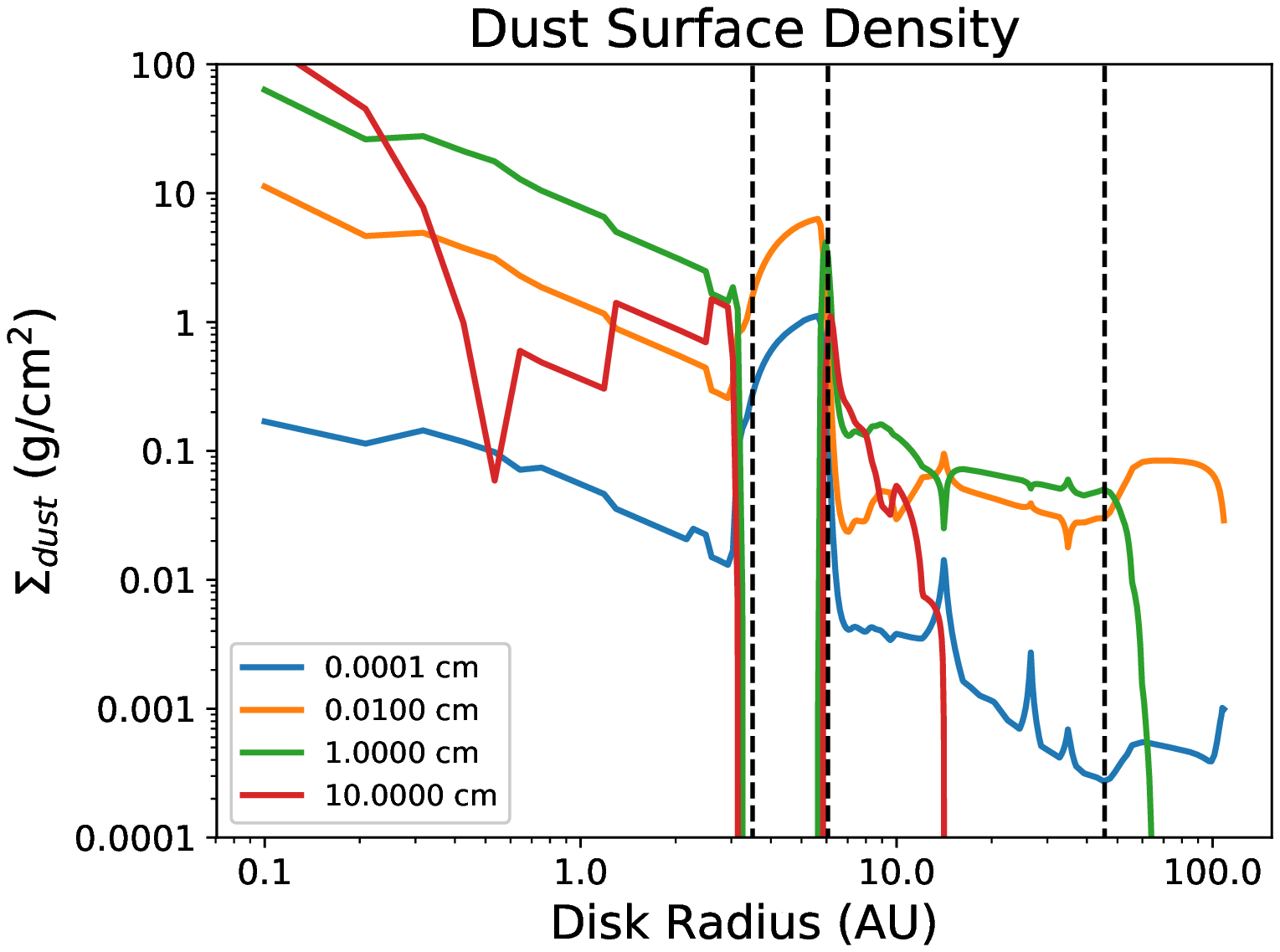}
\put(52,63){ DZ }
\put(64,63){ IL }
\put(78,60){ HT }
\end{overpic}
\caption{Dust surface density radial profile for varying dust sizes at $t_{age} = 0.8 Myr$. We mark the location of the outer dead zone edge (DZ), ice line (IL), and heat transition (HT) with vertical dashed lines, located at $\sim$ 3, 6, and 45 AU respectively. }
\label{fig:glbl01}
\end{figure}

In Figure \ref{fig:glbl01} we show the surface density profile for grains of different sizes. The location of the dead zone edge (3 AU), water ice line (6 AU), and heat transition (45 AU) are marked with vertical dashed lines. Inward of the ice line, the ice sublimates leaving grains that are more susceptible to fragmentation. This destroys the large grain population ($a \gtrsim 1$ cm) as they radially drift across the ice line. 

In the outer region of the disk the size of the dust grains is limited by radial drift, the rate of which is dependent on the  gas pressure gradient (and hence on the gradient of the gas temperature). The temperature profile steepens across the heat transition from larger to smaller radii as the disk heating becomes dominated by viscous evolution, resulting in a slower radial drift rate. We find that this change in the drift rate results in an enhancement for medium sized (a $\sim 0.01$ cm) grains. 

\ignore{
\begin{figure}
\centering
\begin{overpic}[width=0.5\textwidth]{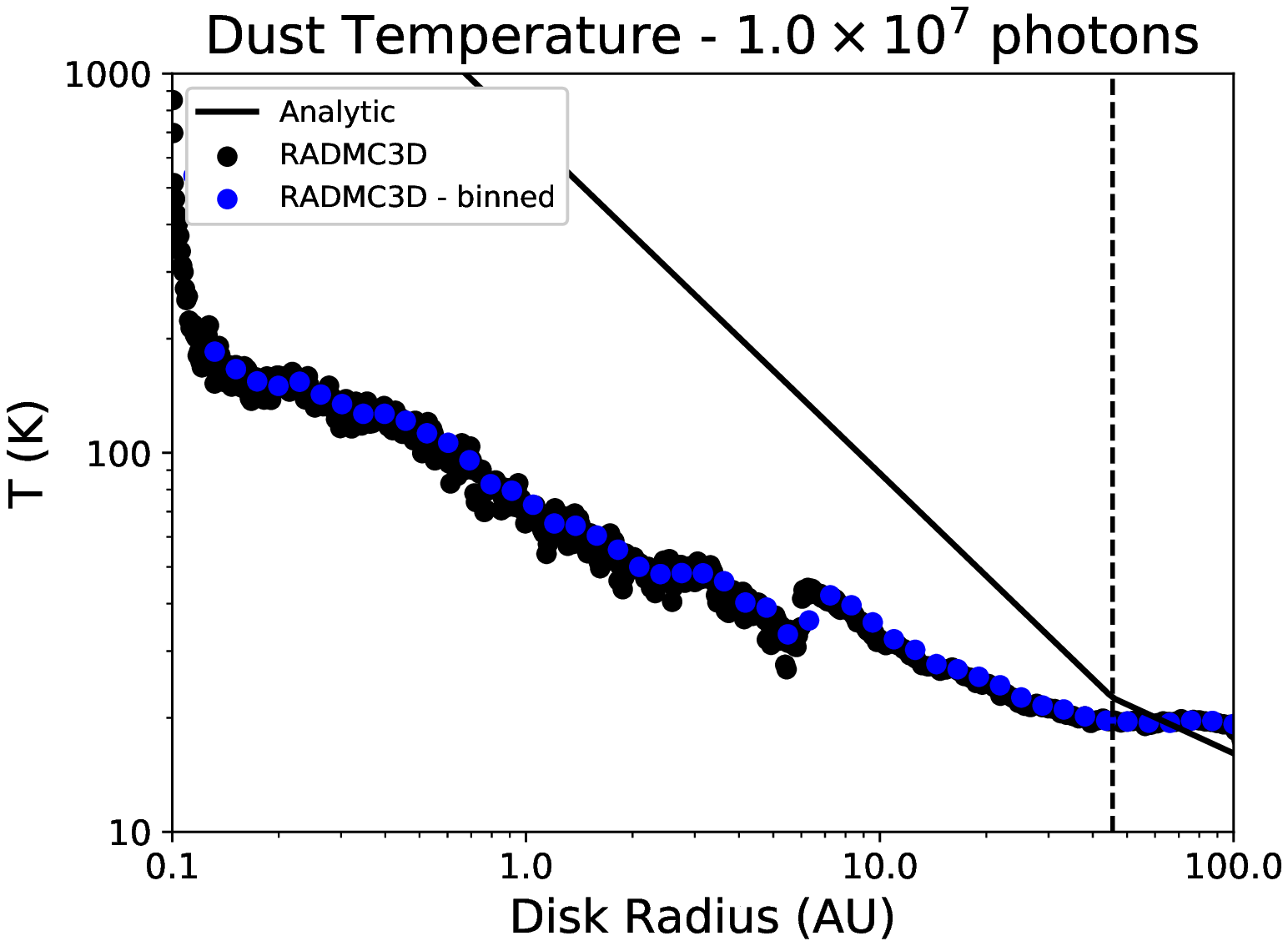}
\put(78,60){HT}
\end{overpic}
\caption{The dust temperature derived by a Monte Carlo radiative transfer run (points) using the dust distribution found in Figure \ref{fig:glbl01}. We similarly binned the data for use in the torque calculation (blue). Included is our analytic temperature model which includes both viscous and radiative heating for reference (solid line). The enhancement of medium sized grains leads to a higher opacity and warmer dust outside the heat transition.}
\label{fig:glbl02}
\end{figure}
}

\begin{figure}
\centering
\subfigure[ The mass and disk radius dependence of the total torque, at a disk age of 0.1 Myr. We note the location of the heat transition (HT) with a vertical dashed line, along with the outer dead zone edge (DZ) and water ice line (IL) features. The colour denotes the magnitude and sign of the total torque, outward migration in blue and inward migration in red. Ultimately the planet opens a gap when it has grown sufficiently large, which shuts off the effect of Type-I migration.\label{fig:glbl04a}]{
\begin{overpic}[width=0.5\textwidth]{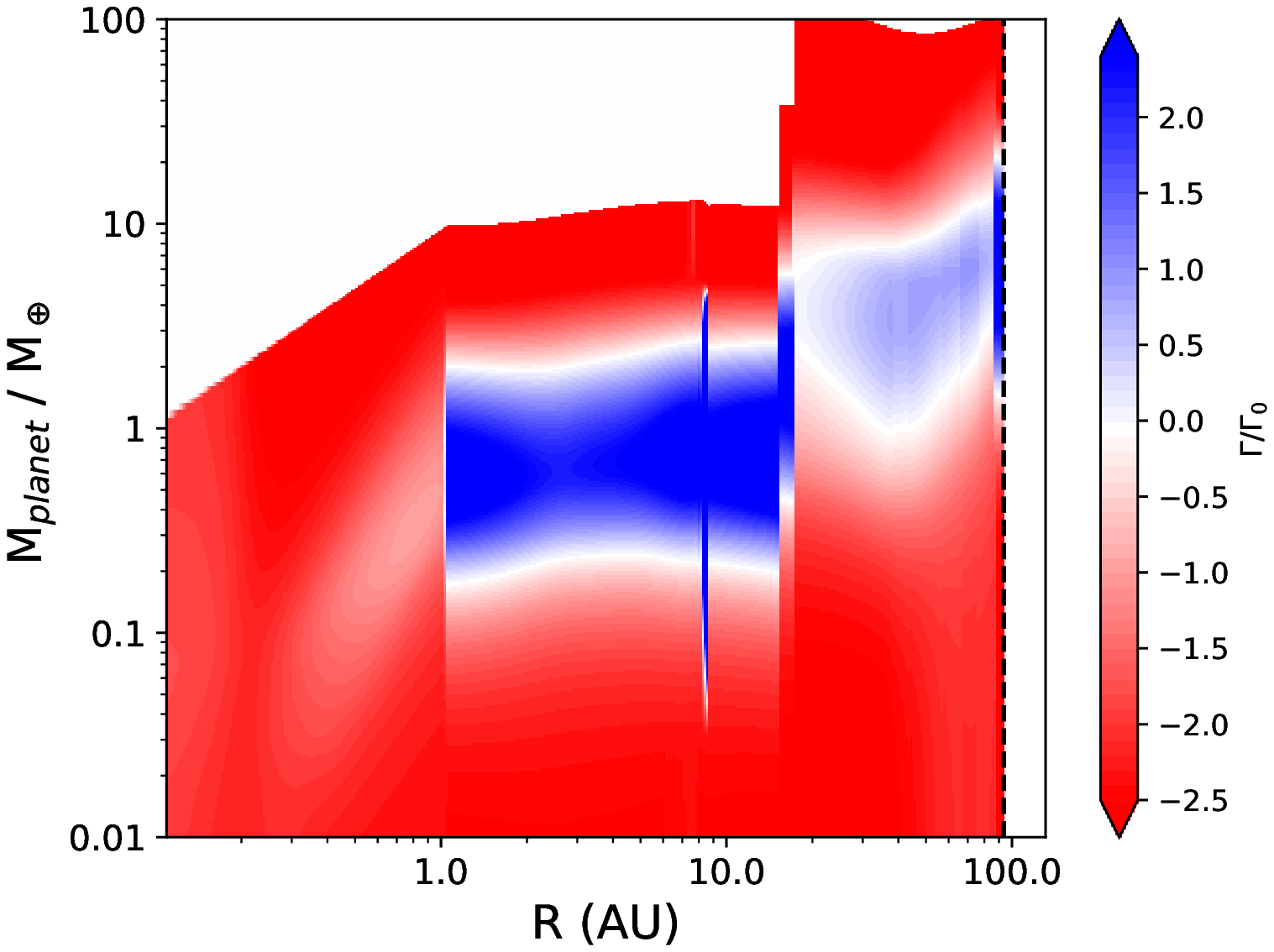}
\put(16,60){ Type-II regime }
\put(49,30){ IL }
\put(61,36){ DZ }
\put(70,65){ HT }
\put(18,15){\Large 0.1 Myr}
\end{overpic}
}
\subfigure[ Same as in Figure \ref{fig:glbl04a}, but for our fiducial disk age of 0.8 Myr. \label{fig:glbl03} ]{
\begin{overpic}[width=0.5\textwidth]{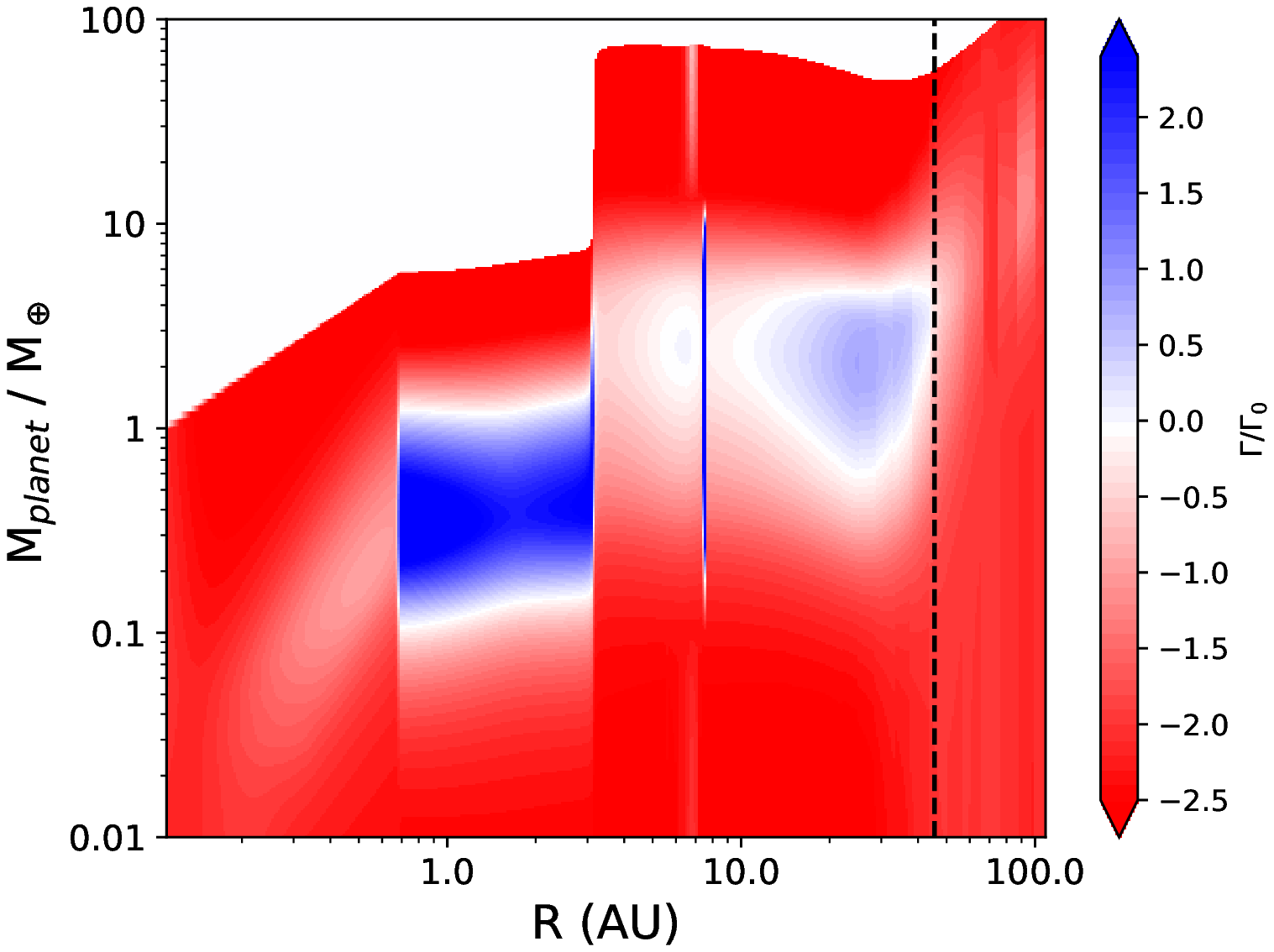}
\put(45,30){ DZ }
\put(50,40){ IL }
\put(65,63){ HT }
\put(18,15){\Large 0.8 Myr}
\end{overpic}
}
\subfigure[ Same as in Figure \ref{fig:glbl04a}, but for an older disk with an age of 1.9 Myr. \label{fig:glbl04b} ]{
\begin{overpic}[width=0.5\textwidth]{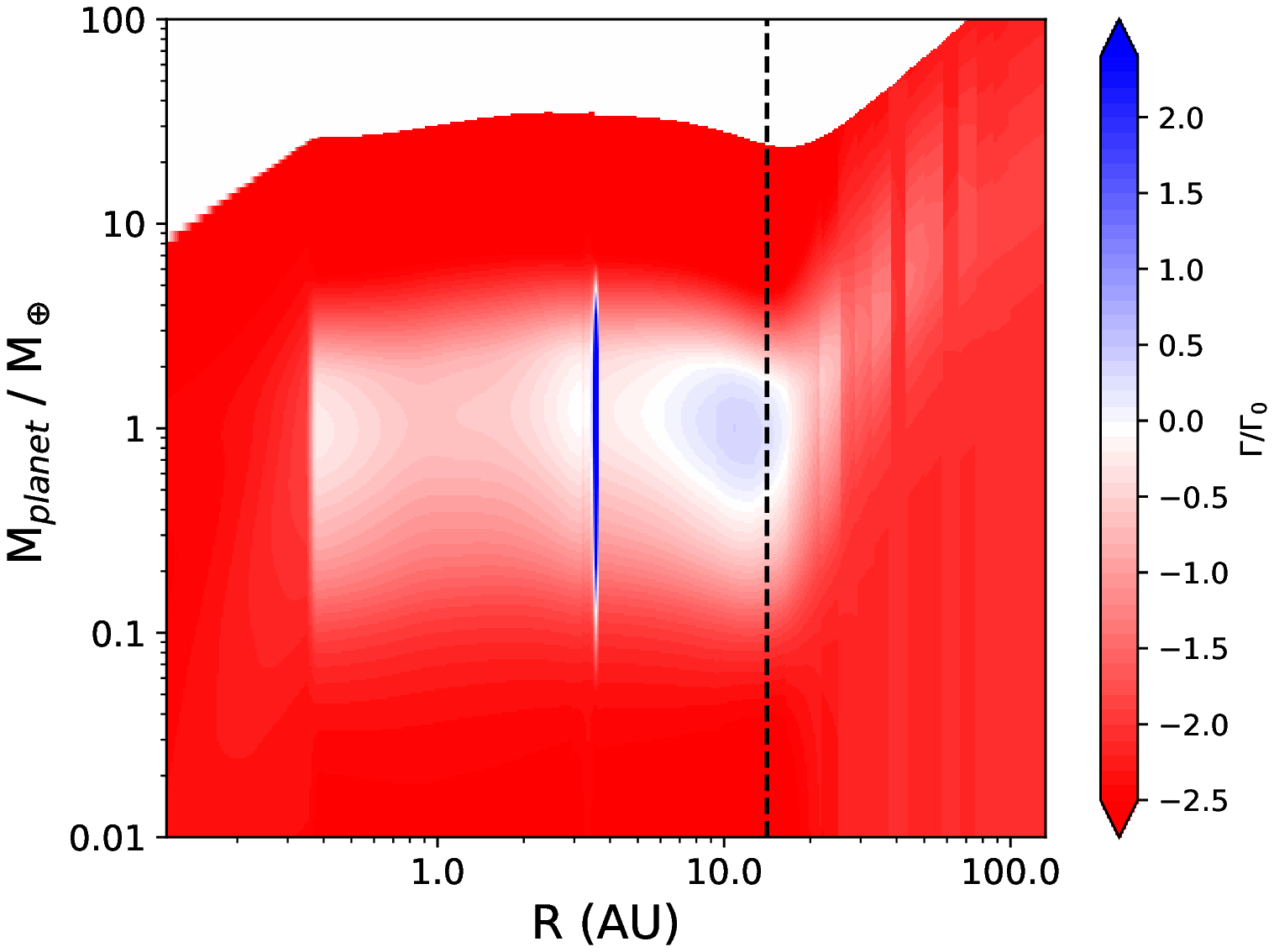}
\put(43,35){ IL } 
\put(52,65){ HT }
\put(18,15){\Large 1.9 Myr}
\end{overpic}
}
\caption{Time evolution of global trapping.}
\label{fig:glbl0304}
\end{figure}

In Figure \ref{fig:glbl0304} we show the time evolution of the total torques in our fiducial model for disk ages of 0.1, 0.8, and 1.9 Myr. We focus on an age of 0.8 Myr shown in Figure \ref{fig:glbl03} as it pertains to the estimated age of the HL Tau system that we have been analyzing up to now. We show the total torques acting on a planet with a given mass at a given position in the disk, up to the gap opening masses in our disk model. Again, in computing the temperature profile of the disk we combine the effects of viscous and radiative heating by taking the sum in equation \ref{eq:Tpro01}, and use our modified model from section \ref{sec:resTrap} to represent the temperature profile in the viscously heated part of the disk. 

We find strong trapping near the water ice line ($\sim$ 6 AU), as well as clear trapping (inward migration outward of the trap and outward migration inward of the trap) at the dead zone edge ($\sim$ 3 AU) for $0.1$ M$_{\oplus}\lesssim$ M$_{plnt}$ $\lesssim$ few M$_{\oplus}$. By inspection the location of the dead zone trap in Figure \ref{fig:glbl03} and the increase in the temperature profile in Figure \ref{fig:glbl02} do not align, which disagrees with our previous assertion for the root cause of planet trapping at the dead zone. Instead it appears that the change of $\alpha_{\rm turb}$ is more important to the net torque. This could be caused by the fact that at low turbulent viscosity (although high enough for the co-rotation torque to remain unsaturated) the co-rotation torque is primarily given by the horseshoe drag, rather than the linear torque \citep{Paard11}. The strength of the horseshoe drag torque has a strong dependence in its entropy component to the temperature gradient, and hence with the steep temperature gradient in our viscously heating regime the horseshoe drag can overpower the outward torque of the Lindblad torque.

Approaching the heat transition from larger radii where the temperature profile flattens, the magnitude of the inward torques are strongly reduced. We find planet trapping at the heat transition for a range of larger masses than is seen for the dead zone and water ice line. This shift could be caused by the higher turbulent viscosity at the lower disk temperatures near the heat transition, which tends to smooth out horseshoe orbits unless the planet is sufficiently massive. This result shifts the mass range where trapping is relevant to slightly higher masses than was assumed in our previous work \citep{HP13,Crid16a,APC16a}. However, the inward migration rate is still an order of magnitude lower than would be expected from standard Type-I migration, for planets with M$<$ 0.5 M$_{\oplus}$. Hence their migration is sufficiently slow that they have enough to time to grow into a mass range where they are effectively trapped.

While a heat transition null point does not appear for low mass planets, nevertheless it will act as an {\it effective} trap. This effective trap reduces the migration rate enough to allow for the growth of a super-earth core, at which point it will be fully trapped. The implications of this effective trapping can be studied through population synthesis models, and is left to future work. One particularly important question is whether a planet migrating due to the torques in Figure \ref{fig:glbl03} will deviate far from a planet that is assumed to be perfectly trapped at the heat transition - since both scenarios result in similar migration timescales (factor of order unity). In our formation model below we assume that the planet is perfectly trapped at the heat transition.

\subsection{Time Evolution of Global Trapping}

We now show how the radial positions of these planet traps change as the disk ages. In our analytic model the disk evolves in two ways: 1) As mass accretes onto the host star the total mass of the disk decreases, reducing both the surface density of the gas and dust at all radii as well as reducing the global mass accretion rate. 2) With the falling mass accretion rate, viscous heating becomes less predominant, lowering the temperature of the gas within the heat transition. Thus the radial position of the heat transition moves inward because the temperature of the gas due to direct irradiation is not susceptible to the same time evolution. With the reduction of the gas and dust surface density, the flux of ionizing photons to the midplane of the disk increases, shrinking the dead zone.

In Figures \ref{fig:glbl04a} and \ref{fig:glbl04b} are formatted the same way as Figure \ref{fig:glbl03}, for an early (0.1 Myr) and late (1.9 Myr) time respectively. At early times the heat transition is not yet present because the disk heating is completely dominated by viscosity. It is also hotter and more dense at these early times which starts the ice line feature (now located at 8 AU) and dead zone edge (now located at 17 AU) farther out than they are observed later on in the disk's lifetime, as seen in Figure \ref{fig:glbl03} at 0.8 Myr. Nevertheless we see similar features at this early time as we saw above. The ice line trap has a mass range for trapping that is shifted to lower mass from what we saw in Figure \ref{fig:glbl03}. This can be attributed to it sitting within the dead zone early on in the disk's life, while the dead zone has crossed it by 0.8 Myr. The dynamical implications for planets trapped at crossing planet traps is an interesting problem that is currently beyond our scope of work.

In Figure \ref{fig:glbl04b} the surface density of the disk has decreased sufficiently that the dead zone is no longer present in the disk, receding towards the host star. Any planets trapped at the dead zone edge would either have been lost, or have grown enough to enter into the slower Type-II migration regime. Indeed in our planet formation calculations we find that the planets formed at the dead zone edge in the high-mass models reach a Type-II migration regime prior to the disappearance of the dead zone. The heat transition, which as moved to a smaller radii at these later times, shows similar {\it effective} trapping features as was observed at 0.8 Myr. Once again we find a region of slower migration appearing near the heat transition point.

\section{ Results: Planets in HL Tau?  }\label{sec:resPlnt4}
We ran each of our disk models described in \S \ref{sec:HLTau} for 2 Myr to account for any possible discrepancy in the observed age of the HL Tau system.  We present the final mass and semi-major axes of formed planets at both 1 and 2 Myr in Table \ref{tab:result01}. A lifetime of 2 Myr is half of the lifetime that we have assumed in our past work \citep{Crid16a,Crid17}, and as a result we found that the dead zone edge was the only planet trap to produce a planet with an appreciable mass in each disk model. The dead zone's rapid shrinking facilitated its success as a position of rapid planet formation. Each of the other traps would have required more time to form a large planet.

For example the planetary embryo trapped at the heat transition, which exists between the first and second gap in the high-mass model of HL Tau only grew by about $10\%$ of its initial mass of 0.01 M$_\oplus$. Such a small planet could not be responsible for the observed dust gaps, and hence it is doubtful that a planet formed through planetesimal accretion could form the observed features near the heat transition. A similar issue would restrict the formation of planets near the CO$_2$ ice line, because it and the heat transition occupy similar regions of the disk. As a result we similarly to not expect large embryos to form at the CO$_2$ ice line.

We note that we have largely ignored the impact of pebble accretion in computing the early evolution of these planetary embryos. \cite{Brouwers2018} show that pebble accretion can efficiently build protoplanetary cores up to a mass of about 0.6 M$_\oplus$ before additional pebbles are ablated in the growing gaseous envelope before reaching the core - stalling growth \citep{Alibert2017}. This stalling becomes less efficient at radii $>$ 20 AU, and hence pebble accretion is expected to be more efficient at building planetary bodies in the outer regions of the disk than planetesimal accretion alone \citep{Bitsch2015}.

Through planetesimal accretion alone it seems unlikely that a sufficiently large planet will grow within the current lifetime of the HL Tau system. Moreover the planetesimal accretion timescale scales inversely with planet mass, so even if we start the embryos with a larger mass, for example the mass attainable by early pebble growth (0.6 M$_\oplus$), we would not expect more efficient growth. Nevertheless, pebble accretion could potentially build Earth-mass objects at the CO$_2$ ice line ($\sim$ 13.6 AU) and heat transition ($\sim$ 43 AU) in the HL Tau system. 

Whether these planets would carve a gap in the dust disk is generally determined through numerical simulations \citep{Duffell2013,Ataiee2018,Bitsch2018} and is beyond the scope of this work. However generally these works find a wide range of planet masses can open an observable (by ALMA) gap in the dust (often called pebble isolation mass). This pebble isolation mass is described by the minimum mass required for a planet to cause a pressure bump outward of its orbital radius, thereby trapping pebbles. \cite{Ataiee2018} derive an empirical relation which describes the minimum mass ratio $q = M_{\rm planet}/M_*$ required to open a dust gap in a disk with scale height $h = H/r$ and turbulent $\alpha$:\begin{align}
\left(\frac{q}{h^3}\right)^2 \approx 82.33\alpha +0.03.
\end{align}
For our disk model we estimate a pebble isolation mass of 0.04 M$_{J} \approx$ 14 M$_\oplus$ at the CO$_2$ ice line (r = 13.6 AU). 

Hence the mass required to open a gap in the HL Tau disk is incompatible with the maximum mass that can be built through planetesimal accretion in the current lifetime of the disk at the CO$_2$ ice line.

\begin{figure}
\centering
\includegraphics[width=0.5\textwidth]{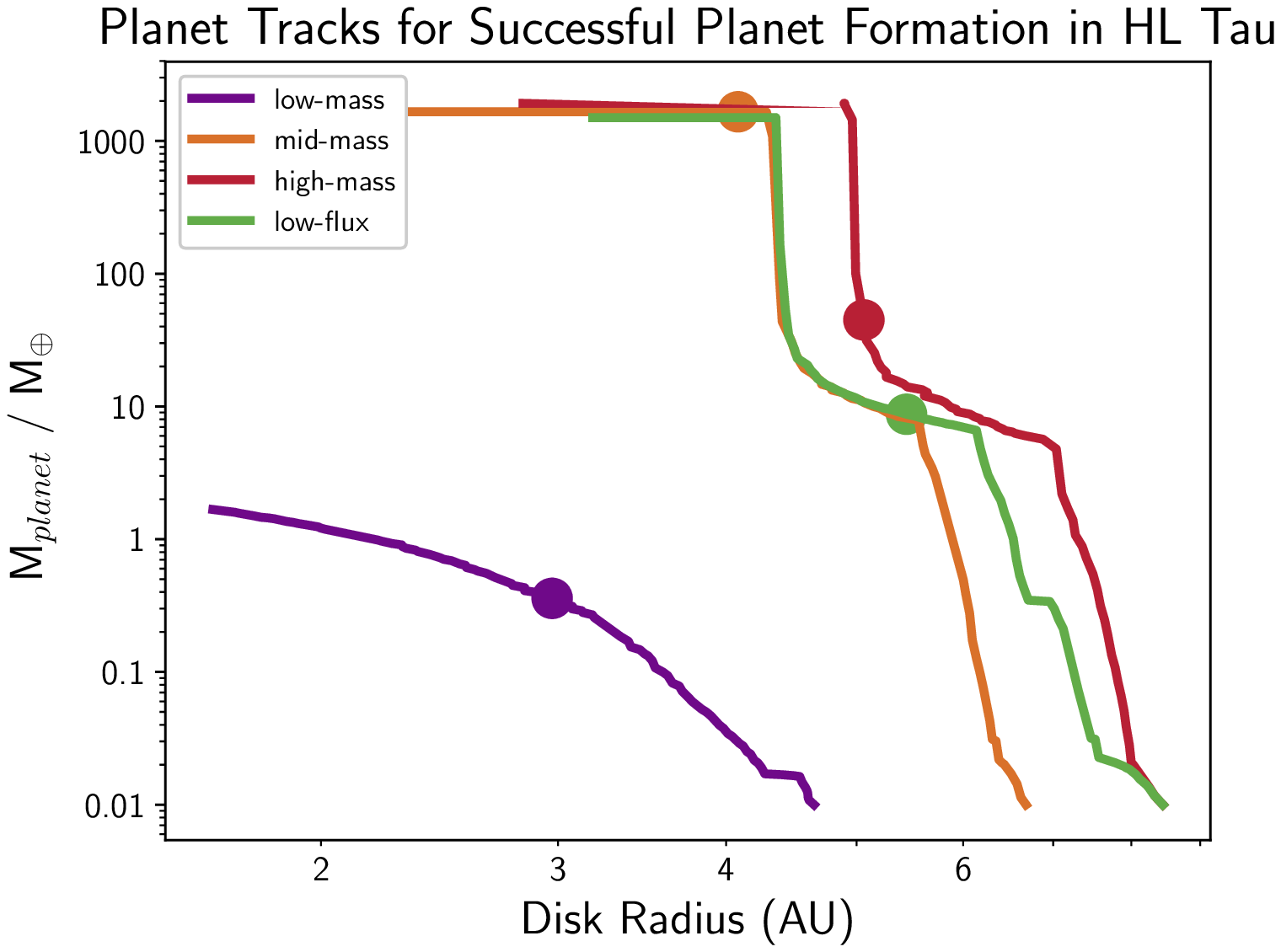}
\caption{ Planet formation tracks for the planets formed at the dead zone trap in the four disk models. The full tracks evolve for 2 Myr, and we note the location of the planet along each track at 1 Myr with a point. }
\label{fig:p4:results401}
\end{figure}

In Figure \ref{fig:p4:results401} we show the `planet tracks' for the successful planets - the evolution through the mass-semi-major axis diagram. Generally speaking the lower disk mass models had smaller dead zones, because their gas surface densities are lower, resulting in a higher flux of radiation at the disk midplane. This leads to the proto-planets beginning their growth at smaller radii in lower mass disks. The high-mass and low-flux model began their formation at the same radii because they had the same initial mass. 

Note that the evolution of the growing planet is not only dictated by the initial mass of its host disk. The low-flux model has a lower UV-flux than the high-mass model, so the low-flux disk tends to be less ionized throughout the evolution of its proto-planet. As a result, the planet growing in the low-flux model migrated inward more slowly, causing it to grow in a lower density region of the disk than the planet growing in the high-mass model. To illustrate this we mark the position of the planet in the mass-semi-major axis diagram at 1 Myr into their evolution with a point. The planet in the high-mass model (red) is at a larger mass and closer to its host star than the planet in the low-flux model (green)- a direct result of the difference in UV-flux.

\begin{table*}
\centering
\caption{Planet orbital properties at 1 and 2 Myr}
\label{tab:result01}
\begin{tabular}{|c|c|c||c|c|}
\hline
Name & M$_{t = 1 Myr}$ (M$_\oplus$) & r$_{t = 1 Myr}$ (AU) & M$_{t = 2 Myr}$ (M$_\oplus$) & r$_{t = 2 Myr}$ (AU) \\\hline
low-mass & 0.36 & 2.2 & 1.67 & 1.3 \\\hline
mid-mass & 1658 & 2.0 & 1658 & 1.5 \\\hline
high-mass & 44.9 & 2.1 & 1919 & 1.3 \\\hline
low-flux & 8.7 & 2.2 & 1500 & 1.1 \\\hline
\end{tabular}
\end{table*}

Marking the location of the forming planets on their formation tracks at the current age of HL Tau (1 Myr), we find that the mid-mass model could produce a large (M $\sim 5.2$ M$_{Jup}$) planet at a radius of about 2 AU. This disk radius is well below the spatial resolution limit of the VLA and ALMA for this source and does not coincide with the orbital radii of the gaps observed by \cite{ALMA15}. \cite{Tamayo15} estimated the radii of the main gaps to be: 13.6, 33.3, 65.1, 77.3, 93.0 AU. None of these gaps coincide with any of our 3 fiducial planet traps (water ice line, dead zone, and heat transition), and hence we cannot account for the dust gaps through our planet formation model alone. We can predict however, that with higher spacial resolution the existence of a young, possibly still forming, proto-planet could be observable inward of the ice line of the HL Tau disk.

\section{ Conclusions }\label{sec:con4}

In this work we have applied a theoretical model for the evolution of the gas surface density, temperature, and chemistry to the HL Tau disk system. We combine this model with the physical processes of planet formation to predict the potential migration and growth of planets in this system. We seek to study the prospects of planet formation in the $\sim$1 Myr old disk as well as explain the gaps that have been observed. 

Due to the uncertainty of the gas mass in the disk we considered three models with differing initial disk masses which, after 1 Myr of evolution, correspond to the range of gas masses inferred for HL Tau. We investigated the question of whether or not dust gaps in disks are caused by proto-planetary embryos trapped at the ice lines of multiple volatiles. We explored the underlying details of planet trapping at these ice lines in both viscously heated disks and found an analytic fit which models the properties of the temperature profile within the ice line. Additionally we explored planet trapping at the CO ice line in the radiatively heated outer parts of the disk. Finally we explore the trapping at other traditional planet traps (dead zone edge and heat transition) to derive a `global' picture of planet trapping in disks like in HL Tau. 

This work focused on answering the following questions:\begin{itemize}
\item Do the radial locations of the dust gaps coincide with volatile ice lines in our disk model?
\item Can these ice lines act as planet traps for young planetary embryos?
\item Can a planet form in this young ($\sim 1$ Myr) disk through the accretion of planetesimals?
\end{itemize}

To answer these questions we have developed a physical theory of planet trapping based on opacity transitions at ice lines, and a detailed physical and chemical structure of protoplanetary disks.

There are several caveats to our assumptions. In particular, in our chemical model we have not accounted for the possibility that volatile species may freeze out onto the ice mantle of a previously frozen out gas species (for example CO freezing onto a mantle of water ice). These different freezing scenarios generally have higher binding energies, and hence higher sublimation temperatures which would shift the ice line location closer to the host star than was reported here. However this shift would be small, since the sublimation temperatures change by only a few Kelvin, and the temperature profile is shallow around the CO ice line. Additionally, our chemical models adopts reaction rates that are based on lab experiments, but does not account for the uncertainties involved in these experiments. Deviations in binding energies could lead to different sublimation temperatures which would again shift the location of ice lines. However we expect these shifts to be small, well within the resolution limit of ALMA. In principle, one could propagate the uncertainties in the binding energies through the astrochemical calculation to determine a range of possible ice line locations.  Another key point is that we present only 3 models whereas an entire distribution of models whose initial conditions are constrained by details of the HL Tau system would
be desirable.  

Nevertheless, our analysis leads to some important general conclusions about planet trapping and gaps:\begin{itemize}
\item We confirm that the opacity transition at the water ice line does lead to planet trapping
\item If the water ice abundance is reduced by at least a factor of 2 relative to the fiducial value of our astrochemical model (1.5$\times 10^{-4}$ per H), trapping is switched off
\item Because of a steeper temperature gradient around its ice line, CO$_2$ is also able to act as a planet trap, even given its underabundance relative to water by a factor of 400
\item Extending to other volatiles, CO is also abundant enough to act as a trap - however it does not - since the opacity transition across its ice line has a gradient that is too shallow
\end{itemize}

\noindent By applying this knowledge to HL Tau, we find that:
\begin{itemize}
\item Our best coincidence between the locations of the CO$_2$ and CH$_4$ ice lines with the inner two dust gaps (13.6 \& 33.3 AU) in the HL Tau disk is found at an age of 0.8 Myr in our high-mass model - favouring a high initial mass in the disk
\item Planet formation through the accretion of planetesimals in the HL Tau disk produces planets with masses up to a 5 M$_{Jupiter}$ within 1 Myr, depending on the initial mass of the disk
\item These planets would orbit in the innermost part of the HL Tau disk at the dead zone trap, located at $ \le 2.5 $ AU, which cannot be currently resolved by ALMA.
\item While the CO$_2$ ice line could trap growing embryos, it exists too far out to efficiently grow a planet through planetesimal accretion
\item If the initial embryo sizes are determined by pebble accretion, then one would expect planets of at least 0.6 M$_\oplus$ located at the ice line traps, but growth from this point would still be limited
\item From simple estimations of the required mass to open a dust gap (pebble isolation mass), planetesimal accretion is not fast enough to grow gap-opening planets at radii $> 10$ AU
\end{itemize}

Generally the ice lines of CO$_2$, CH$_4$, and NH$_3$ are at radii that are too far for planetesimal accretion to build massive planets, but we do expect CO$_2$ to be able to trap small planets, possibly built by pebble accretion. Their coincidence with the location of the gaps in the HL Tau disk remains suggestive of the root cause of these gaps. We have not explored the initial build up of an embryo by a combination of the streaming instability and pebble accretion, and it is possible that larger initial embryos (than our assumed 0.01 M$_\oplus$) are the initial state of planetesimal accretion. We do not expect this to severely alter our results however, since the oligarchic growth timescale scales with M$_{core}^{1/3}$ \citep{IL04a}, the final mass of these embryos will not have changed far from their initial mass.

\section*{ Acknowledgements }

We thank the anonymous referee for their useful comments that greatly improved the manuscript. The work made use of the Shared Hierarchical Academic Research Computating Network (SHARCNET: www.sharcnet.ca) and Compute/Calcul Canada. A.J.C. acknowledges funding from the National Sciences and Engineering Research Council (NSERC) through the Alexander Graham Bell CGS/PGS Doctoral Scholarship. R.E.P. is supported by an NSERC Discovery Grant. M.A. acknowledges funding from NSERC through the PGS-D Alexander Graham Bell scholarship.

\bibliography{mybib}{}
\bsp

\appendix

\section{ Computing refractive indices and opacity transitions }\label{sec:app01}

We follow the work of \cite{Miyake1993} to compute the optical constants of a silicate dust (with size 0.1 $\mu$m) / water ice mixture. We use the Bruggeman effective medium theory which gives a simple combination of the dielectic functions of silicate and ice ($\epsilon_{sil}$ and $\epsilon_{ice}$ respectively) to compute the effective dielectric function ($\epsilon_{eff}$) of the mixture \citep{Miyake1993}:\begin{align}
f_{sil}\frac{\epsilon_{sil}-\epsilon_{eff}}{\epsilon_{sil}+2\epsilon_{eff}} + f_{ice}\frac{\epsilon_{ice}-\epsilon_{eff}}{\epsilon_{ice}+2\epsilon_{eff}} = 0,
\label{eq:MN93}
\end{align}  
where $f_{sil}$ and $f_{ice}$ are volume filling factors: $f_{sil} = \zeta_{sil}\rho_{ice}/(\zeta_{sil}\rho_{ice} + \zeta_{ice}\rho_{sil})$, $f_{ice} = 1 - f_{sil}$. We take the internal densities of silicate grains and ice: $\rho_{sil} = 3.3$ g cm$^{-3}$ and $\rho_{ice} = 0.92$ g cm$^{-3}$. The mass fractions (relative to the gas) $\zeta_{sil}$ and $\zeta_{ice}$ are allowed to vary across the ice line according to the results of our astrochemical model. We compute the dielectric functions $\epsilon_{sil}$ and $\epsilon_{ice}$ using the complex optical constants for silicate \citep{Draine03} and water ice \citep{Warren84} individually. Combining our astrochemical results with equation \ref{eq:MN93}, we can compute the complex spectral indicies ($m_{eff} = n+ik = \epsilon_{eff}^2$) of the ice-grain mixture as a function of radius across the ice line. 

We then compute the radially varying Planck mean opacity of the dust-ice mixture using the internal opacity calculator of the Monte Carlo radiative transfer scheme {\it RAMDC3D} \citep{RADMC} and the stellar temperature of HL Tau (4395 K). Finally with this radially varying opacity we then compute new temperature and gas surface density profiles in a similar manner as is done in our original disk model, assuming that the gas is heated through viscous dissipation (eg. \cite{LB74}). Using these new temperature and surface density profiles we compute the total torques on a planet using the work of \cite{Paard2011} and \cite{Cole14} as a guide. We outline this work below.

\begin{figure}
\includegraphics[width=0.5\textwidth]{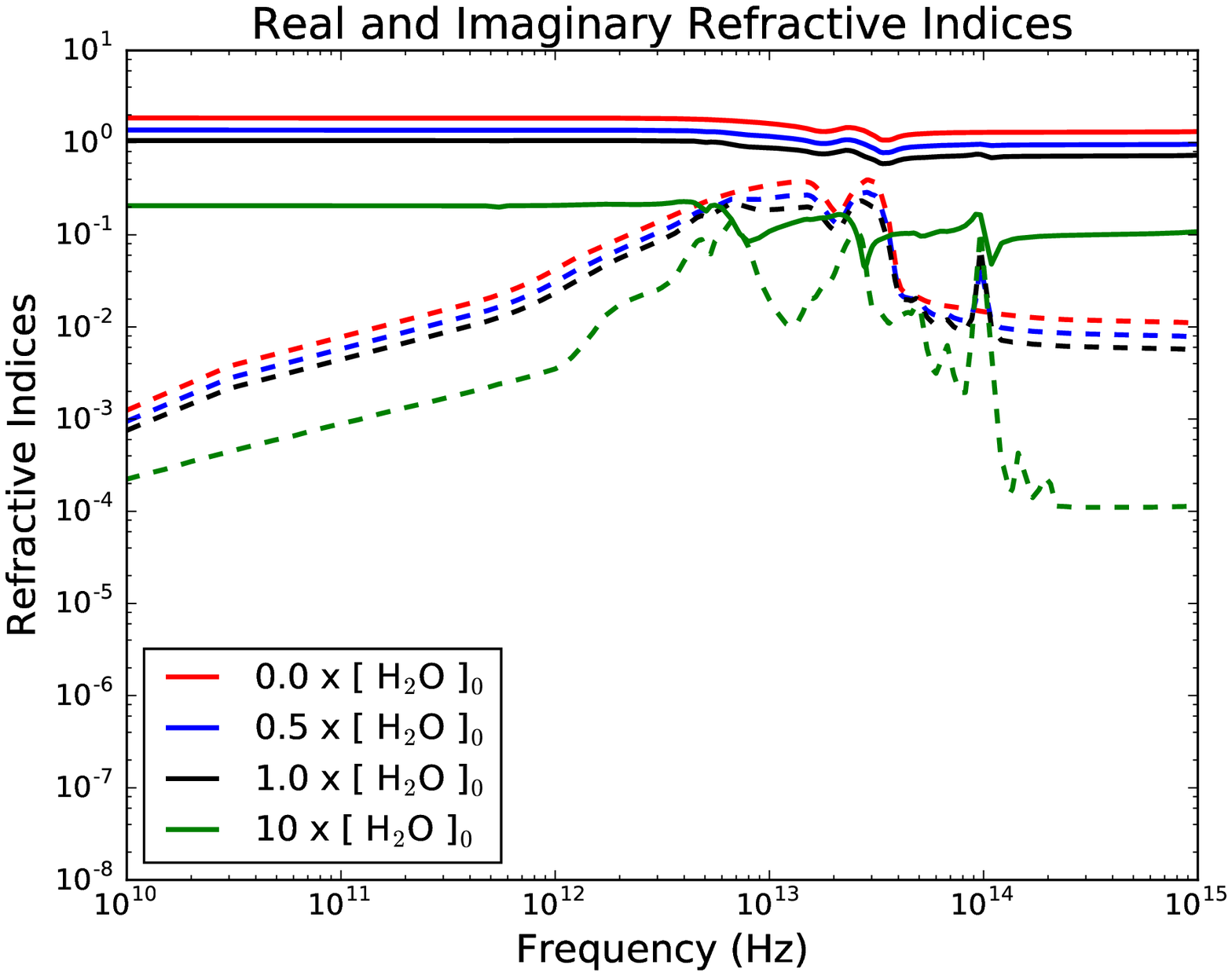}
\caption{Ice mass fraction dependence of the real (solid line) and imaginary (dotted line) components of the complex refractive index of a dust grain / water ice mixture. The abundance of water ice ([ H$_2$O ]$_0$) is set by our astrochemical simulation our disk models. Here we see that across the ice line, as the mass fraction of water ice smoothly increases, the total opacity of the dust / ice mixture will be reduced.}
\label{fig:p4:new01}
\end{figure}

In Figure \ref{fig:p4:new01} we show the dependence of the complex refractive index of a grain / ice mixture on the mass fraction of water ice, following the methods of \cite{Miyake1993}. Because of its dependence on the mass fraction of the ice, we expect that only volatiles that are abundant can contribute to planet trapping. Hence along with the H$_2$O, the CO ice line is likely the only other planet trap that could arise at an ice line in our disk model - we will demonstrate this in detail below. 

To calculate planet trapping at an ice lines we have developed a modified disk model based on \cite{Cham09} which varies the Planck mean opacity as a function of radius across the water ice line, using the complex refractive index featured in Figure \ref{fig:p4:new01} and computed following the method of \cite{Miyake1993}. We used the mass abundance of water ice derived from the astrochemical results of our fiducial disk model as the input for computing the refractive indicies. To derive the opacity of the dust-ice mixture as a function of radius we split the radial profile of water ice abundance into a set of different dust grain populations. Each grain population is differentiated by the abundance of ice that has frozen out onto the grain.

\begin{figure}
\includegraphics[width=0.5\textwidth]{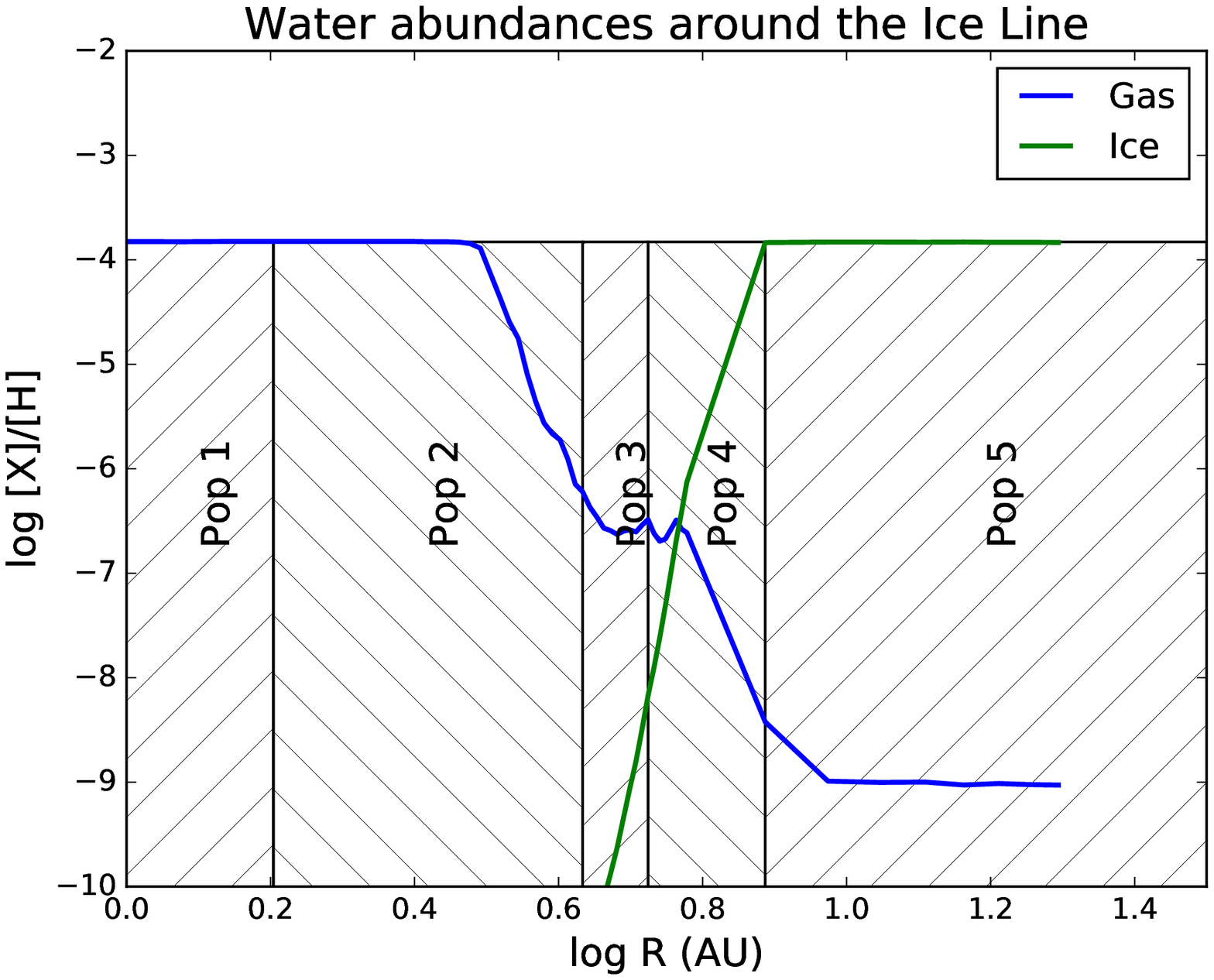}
\caption{An example of splitting the abundances of water ice into 5 grain populations for the purpose of computing dust opacity, evenly distributed over log ice abundance. In the full scale calculation, we use 20 populations that are evenly spaced in log ice abundance.}
\label{fig:p4:new01x}
\end{figure}

In Figure \ref{fig:p4:new01x} we show a simplified calculation based on splitting the ice abundance into different grain populations for the purpose of computing dust opacity. In this work we use 20 populations, equally space in log ice abundance, however for illustrative purposes we show an example with only 5 populations. In the figure, the range of radii where each dust population dominates is shown by the hashed rectangles. Each member of the population is assigned an ice abundance representative of their location in the disk. At radii within the ice line, where the ice abundance changes drastically over only a few AU this method will generally underestimate the average ice abundance of the dust population. For example, Pop 4 in the figure is assigned an ice abundance of $\sim 10^{-8}$ while the majority of the dust will have much higher ice abundances. This inconsistency is reduced by sampling the disk with more dust populations, and in particular populating many dust species inside the ice line. Using 20 dust populations improves our resolution such that it approaches the spacial resolution of our astrochemical results.

\section{ Computing the disk temperature and column density }\label{sec:app01b}

For each dust population we compute the optical constants for a given water ice mass fraction following the method of \cite{Miyake1993}. We then compute the wavelength dependent opacity for each dust population using the internal opacity calculator in {\it RADMC3D}, and finally compute the Planck mean opacity over the radii where each population resides. This last calculation follows: \begin{align}
\kappa(R) = \frac{\int_\lambda \kappa_{i,\lambda} B_\lambda(T(R)) d\lambda}{\int_{\lambda} B_\lambda(T(R))d\lambda},
\label{eq:mod00}
\end{align}
where $B_\lambda$ is the Planck function, $T(R$) is the dust temperature which we first assume is given by the analytic model of \cite{Cham09}, and $\kappa_{i,\lambda}$ is the wavelength dependent opacity of the i'th dust population located at disk radius $R$. This modifies the midplane temperature profile for a viscously heated disk according to \citep{Cham09}:\begin{align}
T^4(R) = \frac{3 \kappa(R)\Sigma(R)T^4_{eff}(R)}{8},
\label{eq:mod01}
\end{align}
where $T^4_{eff} = 9\nu\Sigma\Omega^2/8\sigma$ is the effective temperature of the disk, and $\nu=\alpha c_s^2/\Omega$. The gas surface density is linked to the mass accretion rate through the disk:\begin{align}
\mdot = 3\pi\nu\Sigma = 3\pi\alpha \frac{kT}{\mu m_H} \frac{\Sigma}{\Omega},
\label{eq:mod02}
\end{align} which we assume is constant in space. We use the mass accretion rate and gas surface density determined by our fiducial (high-mass) model to first compute a new temperature profile using equation \ref{eq:mod01} and the radially dependent opacity. Next we compute a new gas surface density profile and radius dependent opacity using the new gas temperature profile and equations \ref{eq:mod02} and \ref{eq:mod00} respectively. With this new gas surface density we recompute the gas temperature profile, then iterate the gas surface density and temperature profiles until they converge. These functions converge in $\sim 5$ repeated calculations of equations \ref{eq:mod00}, \ref{eq:mod01} and \ref{eq:mod02}.

\section{Implementing the Dead Zone into the Two-pop Model}\label{sec:app02}

The Two-population model allows for a radially varying turbulent $\alpha_{turb}$ in computing both the evolution of the two representative grain sizes as well as the reconstruction routines. In our model we smoothly vary $\alpha_{turb}$ such that:\begin{align}
\log_{10}\alpha_{turb} (r,t) = -4 + 2\arctan\left[16(r-r_{dz}(t))\right]/\pi,
\label{eq:fit01}
\end{align}
such that $\alpha_{turb} = 10^{-5}$ within the dead zone and $10^{-3}$ outward of the dead zone edge (located at $r_{dz}$). The position of the dead zone edge evolves as the disk ages. As time passes the disk spreads through viscous evolution, and the large grains drift inward through radial drift. Both effects lead to a reduction in the surface density of dust across the whole disk. The dust is the primary source of opacity to high energy radiation, which dictate the ionization state of the gas, and hence the generation of turbulence through the magnetorotational instability (MRI).

As in our past work we derive the location and evolution of the dead zone edge using the results from our complex photochemical model. In particular, we are interested in the radial distribution of free electrons relative to hydrogen atoms ($x_e$) which evolves in time as the disk ages. From turbulent MHD simulations, a disk is considered turbulently `dead' (ie. $\alpha \lesssim 10^{-5}$) when the growth of the most unstable MRI mode is suppressed by Ohmic diffusion (as well as higher order disspation terms, see below). The impact of Ohmic diffusion is characterized by the Ohmic Elsasser number $\Lambda_O \equiv v^2_{A,z}/\Omega\eta$ where $v_{A,z}$ is the z-component of the Alf\'ven speed, and $\eta$ is the Ohmic resistivity. By assuming that the magnetic field is at equipartition with the fluid then the Alf\'ven speed is related to the gas sound speed: $v^2_{A,z} = 4c_s^2$. The Ohmic resistivity is related to the electron fraction $\eta \propto x_e^{-1}$ (see \cite{Crid16b} for details). We define the dead zone edge where $\Lambda_O = 1$, and ignore higher order sources of dissipation like Ambipolar diffusion and the Hall effect because these are dominated by Ohmic effects on the disk midplane. When we analyse the effect of Ambipolar diffusion in a similar way as Ohmic diffusion (using a separate Elsasser number) we do not find a large difference in the location of the dead zone edge. Analyzing the Hall effect in the simple way presented here is not possible because its impact depends on the orientation of the magnetic field relative to the rotation axis of the disk \citep{BaiStone2016}.
 
Our chemical model produces the evolution of the dead zone edge as a function of time, which we fit to the surface density of the dust and its temperature at the dead zone edge: \begin{align}
r_{dz}(t) = a\Sigma_d^b(r=r_{dz},t)T^c(r=r_{dz},t) {\rm ~AU},
\label{eq:fit02}
\end{align}
We find a weak dependence on the dust surface density ($b = 0.007$), a stronger dependence on its temperature ($c = -1.46$), and an amplitude of $a = 9.93\times 10^3$ for the high-mass disk model. We use the results of this fit to estimate the location of the dead zone edge as a function of time in the Two-population model by solving the equation: \begin{align}
r = a\Sigma_d^b(r,t)T^c(r,t),
\label{eq:fit03}
\end{align}
numerically at every time step. We note a slight inconsistency in using the chemical results of the high-mass model to estimate the location of the dead zone edge. Namely that the chemical results of the original high-mass model had already consisted of a calculation of the dust evolution with the Two-population model without implementing the effect of the dead zone, since we didn't know where the dead zone should show up before running the full model. This is a feature of an `end-to-end' model like ours, where models are run in succession in such a way that their output is used as an input for the next model, while ignoring most back reactions from the models that are farther down the chain.

\section{ Torque Equations }\label{sec:app03}

We follow \cite{Cole14} who write the total torque on a Type-I migrating planet as:\begin{align}
\Gamma_{I,tot} &= \Gamma_{LR} + \left[\Gamma_{VHS}F_{p_\nu}G_{p_\nu} + \Gamma_{EHS}F_{p_\nu}F_{p_\chi}\sqrt{G_{p_\nu}G_{p_\chi}} \right. \nonumber\\
&+ \left. \Gamma_{LVCT}\left(1-K_{p_\nu}\right) + \Gamma_{LECT}\sqrt{\left(1-K_{p_\nu}\right)\left(1-K_{p_\chi}\right)}\right],
\end{align}
where $\Gamma_{LR}$, $\Gamma_{VHS}$, $\Gamma_{EHS}$, $\Gamma_{LVCT}$, and $\Gamma_{LECT}$ are the Lindblad torques, vorticity and entropy-related horseshoe drag torques, and linear vorticity and entropy-related corotation torques, respectively, and are given in equations 3-7 in \cite{Paard2011}. The damping functions $F_{p_\nu}$, $F_{p_\chi}$, $G_{p_\nu}$, $G_{p_\chi}$, $K_{p_\nu}$, and $K_{p_\chi}$ are related to the ratio of either the viscous/thermal diffusion timescales with the horseshoe libration/horseshoe U-turn timescales. They are given by equations 23, 30, and 31 in \cite{Paard2011}. Note that we have neglected the extra damping functions $F_L$, $F_e$, and $F_i$ computed by \cite{Cole14} which depend on the eccentricity and inclination of the planet because we assume that our planets are in circular orbits with zero inclination.

\label{lastpage}

\end{document}